\documentclass{amsart}

\usepackage{enumitem}

\usepackage{amsfonts,amsbsy,amssymb,amsmath,amsthm,amsfonts,mathtools}
\usepackage{color}
\usepackage{xcolor}
\usepackage{mathrsfs}
\usepackage{tikz}
\usetikzlibrary{positioning}
\usepackage{subfigure}
\usepackage{float}

\usepackage{hyperref}
\hypersetup{
    colorlinks=true,
    linkcolor=blue,
    citecolor=blue,
    urlcolor=blue}

\newcommand{\Ann}{\operatorname{Ann}}
\newcommand{\coeffId}[1]{\left[#1\right]_{1_G}}
\newcommand{\Tr}{{\mathrm{Tr}}}

\newtheorem{theorem}{Theorem}[section]
\newtheorem{lemma}[theorem]{Lemma}
\newtheorem{prop}[theorem]{Proposition}
\newtheorem{cor}[theorem]{Corollary}

\theoremstyle{definition}
\newtheorem{definition}[theorem]{Definition}
\newtheorem{example}[theorem]{Example}

\theoremstyle{remark}
\newtheorem{remark}[theorem]{Remark}

\numberwithin{equation}{section}

\begin{document}

\title[Duality on group algebras \& additive group codes]{Duality on group algebras over  finite chain rings: applications to additive group codes
 }


\author[M. Bajalan]{Maryam Bajalan}
\address{Institute of Mathematics and Informatics, Bulgarian Academy of Sciences, Bl. 8, Acad. G. Bonchev Str., 1113, Sofia, Bulgaria}
\curraddr{}
\email{maryam.bajalan@math.bas.bg}
\thanks{The work of M. Bajalan was supported by the Bulgarian MES, Grant No. D01-98/26.06.2025 for NCHDC–part of the Bulgarian National Roadmap on RIs.}

\author[J. de la Cruz]{Javier de la Cruz}
\address{Department of Mathematics, Universidad del Norte, Barranquilla, Colombia}
\curraddr{}
\email{jdelacruz@uninorte.edu.co}
\thanks{J. de la Cruz gratefully acknowledges partial support from the German Academic Exchange Service (DAAD) through a re‑invitation grant during his visit to the University of Bayreuth, Germany.}

\author[A. Fotue]{Alexandre Fotue Tabue}
\address{Department of Mathematics, University of Bertoua, Cameroon}
\curraddr{}
\email{alexfotue@gmail.com}
\thanks{A. Fotue is supported by a "Research in Pairs" grant by CIMPA while his visit to the Institute of
Mathematics, University of Valladolid, Spain.}
\author[E. Mart\'inez-Moro]{Edgar Mart\'inez-Moro}
\address{Institute of Mathematics, University of Valladolid, Castilla, Spain}
\curraddr{}
\email{edgar.martinez@uva.es}
\thanks{E. Martínez-Moro is partially supported by the Grant PID2022-138906NB-C21 funded by MICIU/AEI/10.13039/501100011033 and by ERDF/EU}


\date{\today}

\dedicatory{
}

\begin{abstract}   
Given a finite group $G$ and an extension of finite chain rings $S|R$, one can consider the group rings $\mathscr{S} = S[G]$ and $\mathscr{R} = R[G]$.  
The group ring $\mathscr{S}$ can be viewed as an $R$-bimodule, and any of its $R$-submodules naturally inherits an $R$-bimodule structure; in the framework of coding theory, these are called \emph{additive group codes}, more precisely a (left) additive group code of is a linear code which is the image of a (left) ideal of
a group algebra via an isomorphism which maps $G$ to the standard basis of $S^n$, where $n=|G|$.  
In the first part of the paper,  the ring extension $S|R$ is studied, and several  $R$-module isomorphisms are established for decomposing group rings, thereby providing a characterization of the structure of additive group codes. In the second part, we construct a symmetric, nondegenerate trace–Euclidean inner product on $\mathscr{S}$.
Two additive group codes   $\mathcal{C}$ and $\mathcal{D}$ form an \emph{additive complementary pair} (ACP) if $\mathcal{C} + \mathcal{D} = \mathscr{S}$ and $\mathcal{C} \cap \mathcal{D} = \{0\}$.   For two-sided ACPs, we prove that the orthogonal complement of one code under the trace–Euclidean duality is precisely the image of the other under an involutive anti-automorphism of $\mathscr{S}$, linking coding-theoretical ACPs with module orthogonal direct-sum decompositions, representation theory, and the structure of group algebras over finite chain rings.
 
\end{abstract}

\maketitle
\tableofcontents
\section{Introduction}\label{sec1}

 Error‐correcting codes over finite rings were introduced in 1963 in the paper \cite{AM63}. Still, they did not gain significant interest in the coding theory community until 1994, when the celebrated paper \cite{Z4} showed that the families of nonlinear binary codes --the Kerdock and Preparata codes-- admit a linear structure when they are seen as codes over the ring $\mathbb{Z}_4$. This fact has promoted a significant avenue of research into linear codes over finite rings, which continues to this day. One must notice that it was proved in \cite{Wood} that the most general class of rings suitable as an alphabet for a linear code on which many of the core theorems of classical coding theory hold -- especially those dealing with duality-- is the class of Frobenius rings. For a Frobenius ring $R$, one defines a code over $R$ as any subset of $R^n$, and we say that the code is linear when it forms a submodule of $R^n$. 
If we broaden the notion of linearity, one can think of additive codes where the alphabet is just an arbitrary finite abelian group. They were related at the very beginning with association schemes \cite{Del}. Still, recently, additive codes have assumed greater significance owing to their applications in the construction of quantum error‐correcting codes by using the stabilizer formalism, i.e., building a
bridge between dual-containing additive codes and quantum stabilizer codes \cite{quan1}. Note that additive cyclic codes over fields have been widely studied, see for example \cite{Bierbrauer, CyclicBierbrauer, EncBierbrauer} and the references therein.
When defined over extensions of finite chain rings -- particularly those arising from Galois or Eisenstein constructions-- these codes exhibit deep interactions between ring theory, module theory, and coding-theoretic properties, leading to new insights \cite{JS24c, LS24}.

Finite chain rings are a family of local rings characterized by their unique chains of ideals, which provide a natural setting for code construction due to their modular arithmetic properties and associated duality theories. They are the first and most widely used example of Frobenius rings in coding theory. Seminal works by \cite{CL73} and \cite{GM73} established foundational results on the structure and enumeration of finite chain rings, highlighting their role as extensions of Galois rings. These rings admit two principal types of extensions: \textit{Galois extensions}, which preserve the residue field and are unramified, and \textit{Eisenstein extensions}, which are ramified and constructed via Eisenstein polynomials \cite{CL73, GM73}. Recent studies by \cite{JS25, JS24b} have further explored additive cyclic codes over such rings, emphasizing their duality properties and enumeration. However, the study of \emph{group codes}-- additive codes invariant under group actions-- over extensions of chain rings remains comparatively underdeveloped, despite their potential for symmetry-driven optimization.

The concept of \textit{complementary pairs} of codes, particularly Linear Complementary Pairs (LCPs) and their additive analogues (ACPs), has gained prominence due to applications in cryptography, such as fault-tolerant computation and side-channel attack resistance \cite{GMS20}. A pair of codes $(C, D)$ forms an ACP if $C \oplus D = R^n$ (for ambient space $R^n$) and both are additive. Such pairs ensure that errors correctable by one code are detectable by the other, enhancing security in dual-code schemes.  LCPs and ACPs of codes have been extensively studied over fields; see, for example, \cite{SYS24a, BCW,Choi2023, SHI2022} and the references therein.  LCD codes have also been studied over rings; see, for example, \cite{JS25,Liu2,Liu1}, as well as for additive cyclic codes \cite{JS24b}. Still, their extension to additive codes over rings in more general group-invariant settings presents unique challenges and opportunities. Recent advances include \cite{GMS20}, who characterized LCPs of group codes over finite chain rings, and \cite{BDTM25}, who explored trace duality for additive cyclic codes. Nevertheless, a comprehensive theory for ACPs of group codes over \emph{extensions} of chain rings -- incorporating both Galois and Eisenstein structures -- has been lacking.

This paper bridges this gap by studying, from a module-theoretical perspective,  \emph{additive group codes} over extensions of finite chain rings and their \emph{additive complementary pairs (ACPs)}. Our work unifies Galois and Eisenstein extensions, which are usually treated independently, under a single framework, namely \emph{Galois-Eisenstein extensions}, and establishes their ubiquity (Theorem~\ref{thm:EG}). The key contributions of the work include:
\begin{enumerate}
    \item \emph{Structural Decomposition}. For a finite abelian group $G$ acting on codes, we decompose the group ring $\mathscr{S} = S[G]$ (where $S|R$ is a Galois-Eisenstein extension) into primitive central idempotent elements in Corollary~\ref{cor:idempotent-S}, extending results of \cite{MOO18} to non-cyclic groups. This decomposition classifies left additive $G$-codes as direct sums of ideals in component rings (see Theorem~\ref{thm3.10}).
    
    \item \emph{Duality Theory}. We introduce a \emph{trace-Euclidean-like inner product} $\circledast$ in Section~\ref{ssec:dual}, leveraging the trace map of Galois extensions and the $\gamma$-adic decomposition of Eisenstein extensions. This leads to an explicit characterization of the dual $C^{\perp_{\circledast}}$ for additive abelian group codes (Corollary   ~\ref{cor-dual-abelian}), generalizing the duality results of \cite{BDTM25} to group rings.
    
    \item \emph{ACP Characterization}. Given the involutive ring anti-automorphism $\mu: \mathscr{S} \to \mathscr{S}$, defined by $\mu\left(\sum_{g\in G} a_{g} g\right) = \sum_{g\in G} a_{g} g^{-1}$, we prove that for a pair $\{C, D\}$ of two-sided additive $G$-codes, $\mu(C) = D^{\perp_\circledast}$ holds if and only if $\{C, D\}$ is an ACP (Theorem~\ref{thm:charpair}) in the case the characteristic of the ring is coprime with $|G|$. This extends \cite[Lemma 2.5]{GMS20} to mixed alphabets and provides a duality-preserving isomorphism via the $\mathscr{R}\overline{\mathscr{R}}$-ary image $\Theta$ (Theorem \ref{thm4:main}). For $G$ being an abelian group, we derive enumerative conditions for ACP existence (Proposition~\ref{prop:AbelianACP}).    
\end{enumerate}

 Our results unify and generalize previous work on LCPs~\cite{GMS20}, trace duality~\cite{BDTM25}, and additive cyclic codes~\cite{JS24b} on a standard module theoretical setting.  
The structure of the paper is as follows. Section~\ref{sec:extensions} is devoted to the study of extensions of finite chain rings and the trace map associated with Galois extensions. We characterize all such extensions in Theorem~\ref{thm:EG}. In Section~\ref{sec:gR}, we present the main definitions and properties of group rings over chain rings and introduce additive codes over a chain-ring extension. We treat in detail the special case where $G$ is an Abelian group in Subsection~\ref{ssec:CRT}. Section~\ref{sec:dual} addresses the duality of left additive group codes and their representation as mixed-alphabet codes; several results are specialized to the Abelian-group case. Finally, in Section~\ref{sec:acp}, we investigate additive complementary pairs (ACPs) of group codes over finite chain rings. We provide a complete characterization in terms of duality and an involutive map, and we show that ACPs over Galois–Eisenstein rings descend to ACPs over their residue fields, which facilitates code construction. Illustrative examples over $\mathbb{Z}_9[\alpha,\gamma]$ with symmetric group actions are provided.

\section{Extensions of finite chain rings}\label{sec:extensions}
A \emph{finite chain ring} is a finite commutative ring whose lattice of ideals is totally ordered under inclusion. The following proposition presents several equivalent ways to characterize a chain ring.
\begin{prop}[Proposition 2.1 in \cite{dinh2004cyclic}] For a finite commutative ring $R$, the following conditions are equivalent:
\begin{enumerate}
    \item $R$ is a local ring and the maximal ideal of $ R$ is principal;
    \item $R$ is a local principal ideal ring;
    \item $R$ is a chain ring.
\end{enumerate}
\end{prop}

Throughout this paper, we let $R$\label{R} and $S$\label{S} denote two finite chain rings with respective identity elements $1_R$ and $1_S$. Let $\mathfrak{m}_R$ and $\mathfrak{m}_S$ denote their unique maximal ideals, and let $R^\times$ and $S^\times$ be the respective groups of units. Let $s_1$ and $s_2$ be the nilpotency indices of $\mathfrak{m}_R$ and $\mathfrak{m}_S$, i.e., the smallest positive integers such that $\mathfrak{m}_R^{s_1} = 0$ and $\mathfrak{m}_S^{s_2} = 0$, respectively. We fix the characteristic of $R$ to be $p^e$, where $p$ is a prime number. We denote by $\mathbb{F}_q$ the finite field with $q$ elements, and it is the residue field $R / \mathfrak{m}_R$ when $q = p^d$, and we denote by $\pi \colon R \to \mathbb{F}_q$\label{pi} denote the natural projection.

We say that $S$ is a \emph{ring extension} of $R$, denoted $S|R$, if $R$ is a subring of $S$ and $1_R = 1_S$. We will say that the extension  $S|R$  is an \emph{extension of finite chain rings} if both $S$ and $R$ are finite chain rings.  If $S|R$ is an extension of finite chain rings, then it satisfies the relations $S \mathfrak{m}_R \subseteq \mathfrak{m}_S$ and $\mathfrak{m}_R = \mathfrak{m}_S \cap R$. In this work, we focus on two particular types of ring extensions of a finite chain ring, namely \textit{Galois extensions} and \textit{Eisenstein extensions}.
\subsection{Galois extensions of finite chain rings}
A monic polynomial $f(x) \in R[x]$\label{f(x)} of degree $\ell$ is said to be \emph{basic-irreducible} over $R$ if $\pi(f(x))$ is irreducible over $\mathbb{F}_q$. A ring extension $S|R$ is called a \emph{Galois extension of degree $\ell$} if $S = R[\alpha]$, where $\alpha = x + \langle f(x) \rangle$ and $f(x)$ is a monic basic-irreducible polynomial of degree $\ell$ over $R$.
Such a Galois extension $S$ of $R$ of degree $\ell$ is unique up to $R$-algebra isomorphism (see \cite[Theorem 5.8]{GM73}). A ring extension $S|R$\label{R[alpha]} is said to be \emph{unramified} if $\mathfrak{m}_S = S\mathfrak{m}_R$. Also, it is a well-known fact that the extension $S|R$ is Galois if and only if it is unramified (see \cite[Theorem 5.3]{GM73}).
In particular, a Galois extension of $\mathbb{Z}_{p^e}$ of degree $d$ is referred to as a \emph{Galois ring of rank $d$ over $\mathbb{Z}_{p^e}$}, and we denote it by $\mathrm{GR}(p^e, d)$. It is well known that the characteristic of any finite chain ring $R$ is a power of a prime, say $p^e$. Therefore, $R$ is an extension of the Galois ring $\mathrm{GR}(p^e, d)$\label{GR}, and its residue field is $\mathbb{F}_{p^d}$ (see \cite[Section 1]{CL73}). In this case, the Galois ring $\mathrm{GR}(p^e, d)$ is referred to as the \emph{coefficient ring} of $R$.
It is important to note that every monic polynomial that is basic-irreducible over the coefficient ring of $R$ is also basic-irreducible over $R$ itself. 

The \emph{Teichmüller set} of the ring $R$, denoted by $\Gamma(R)$, is defined as the set of roots 
of the polynomial $x^{p^d} - x$ in the coefficient ring $\mathrm{GR}(p^e, d)$ of $R$. 
More precisely,
\begin{equation*}
    \Gamma(R) = \{ x \in \mathrm{GR}(p^e, d) \mid x^{p^d} = x \}.
\end{equation*}\label{Ga(R)}
We denote by $\Gamma(R)^* = \Gamma(R) \setminus \{0_R\}$ the set of non-zero elements in $\Gamma(R)$. The set $\Gamma(R)^*$\label{Ga(R)*} forms a subgroup of $R^\times$ that is isomorphic to $\mathbb{F}_{p^d}^\times$, where $R^\times$ and $ \mathbb{F}_{p^d}^\times$ denote the multiplicative groups of units of $R$ and $\mathbb{F}_{p^d}$, respectively.

\begin{remark}\label{re-ex} Let $R$ and $S$ be two finite chain rings with coefficient rings $\mathrm{GR}(p^{e}, d)$ and $\mathrm{GR}(p^{e}, r)$, respectively.  
If $S$ is a Galois extension of $R$, then $d$ divides $r$.  
In particular, $\mathrm{GR}(p^{e}, r)$ is a Galois extension of $\mathrm{GR}(p^{e}, d)$ if and only if $d$ divides $r$ (see \cite[Theorem 1.6]{GM73}).
\end{remark}
 
\begin{remark}\label{m22}
Let $R$ be a finite commutative chain ring and $f(x)\in R[x]$ be a monic basic-irreducible polynomial with the root $\alpha$. Then, the Galois extension $S|R$ where $S=R[\alpha]$ is a chain ring, see \cite[Lemma 3.1]{dinh2004cyclic}. Note also that  $\mathfrak{m}_S  = S \mathfrak{m}_R$, thus if  $\gamma$ is the generator $\mathfrak{m}_R$, then
$\mathfrak{m}_S = S\langle \gamma \rangle_R = \langle \gamma \rangle_R$. 
\end{remark}

For each $y \in R$, there exists a unique \emph{Teichmüller representative} 
$x \in \Gamma(R)$ such that $y - x \in\mathfrak{m}_R$. If we fix an element  $\gamma_1 \in \mathfrak{m}_R \setminus \mathfrak{m}_R^2$, then every element of $R$ has a unique $\gamma_1$-\emph{adic decomposition}. In other words,  for each $x \in R$, there exists a unique tuple $(x_0, \ldots, x_{s_1-1}) \in \Gamma(R)^{s_1}$ 
such that 
$$
x =  x_0 + x_1\gamma_1 + \cdots + x_{s_1-1}\gamma_1^{s_1-1}.
$$

From the above reasoning, it follows that $R^\times=\Gamma(R)^*(1+\mathfrak{m}_R)$. Furthermore, if $S = R[\alpha]$, where $\alpha$ is a root of a monic basic-irreducible polynomial of degree $\ell$ over $R$, then $S$ is a free $R$-module with an ordered  basis
$
\underline{\alpha} = (1, \alpha,\alpha^2, \dots, \alpha^{\ell-1}),
$\label{basis-alpha}
and for any arbitrary ordered basis $\underline{\beta} = (\beta_0, \beta_1, \dots, \beta_{\ell-1})$ \label{basis-beta} there exists a non-singular matrix $\mathrm{P}$ over $R$ such that 
$\underline{\beta}=(1, \alpha, \cdots, \alpha^{\ell-1})\mathrm{P}$ (see \cite[Theorem 4.3]{GM73}).
Moreover, if $\underline{\beta} = (\beta_0, \beta_1, \dots, \beta_{\ell-1})$ is a basis of the free $R$-module $S$, then the map $\sigma_q :S\rightarrow S$\label{sigmaq} defined as  $$\sigma_q\left(\sum\limits_{i=0}^{\ell-1}a_i\beta_i\right)=\sum\limits_{i=0}^{\ell-1}a_i\beta_i^{q}$$ generates the group $\mathrm{Aut}_R(S)$ of $R$-linear ring automorphisms of $S$ that fix $R$ pointwise. The \textit{trace map} of the Galois extension $S | R$ is the $R$-module epimorphism
$\Tr : S \longrightarrow R$, \label{trace_map} defined as 
$$
\Tr = \sum_{i=0}^{\ell-1} \sigma_q^i.
$$ 
It is straightforward that $\Tr(1) = \sum_{i=1}^\ell \sigma^i_q(1) = \sum_{i=1}^\ell 1 = \ell$. 
\begin{lemma}\label{non-degenerate}
The symmetry bilinear form $\mathrm{T}_R^S \colon S \times S \to R$\label{trace_bil} defined by $$\mathrm{T}_R^S(x, y) = \Tr(xy)$$ is non-degenerate.
\end{lemma}

\begin{proof}
Let $K = R / \mathfrak{m}_R \cong \mathbb{F}_q$ and $L = S / \mathfrak{m}_S$ be the residue fields of the rings $R$ and $S$, respectively.  Consider the natural projection $\pi: S \to L$.
Since the ring extension $S | R$ is unramified, the field extension $L | K$ is Galois of degree $\ell$. It is well-known that the standard trace map $\operatorname{tr}_{L/K}:  L\to K$  is non-degenerate. Now suppose there exists an element $x \in S$ such that $\mathrm{T}_R^S(xy) = 0$ for all $y \in S$.  To show that $\mathrm{T}_R^S$ is non-degenerate, it suffices to prove that $x = 0$. We obtain
$$
0=\pi(\Tr(xy)) = \operatorname{tr}_{L/K}(\pi(xy))=\operatorname{tr}_{L/K}(\pi(x) \pi(y)).
$$  
Since $\pi$ is surjective, the set $\{ \pi(y) \mid y \in S \}$ spans all of $L$, and hence $\operatorname{tr}_{L/K}(\pi(x) z) = 0$ for all $z \in L$. By the non-degeneracy of $\operatorname{tr}_{L/K}$, it follows that $\pi(x) = 0$, i.e., $x \in \mathfrak{m}_S$. Let $\mathfrak{m}_R=\gamma_1R$. As $S | R$ is an unramified extension, $\mathfrak{m}_S=\gamma_1S$. So $x = \gamma_1 x_1$ for some $x_1\in S$.
Then $0 = \Tr(xy) = \Tr(\gamma_1 x_1 y) = \gamma_1 \Tr(x_1 y)$ for all $y$, hence $\Tr(x_1 y) \in \mathfrak m_R$ for all $y$, that is $\pi(\Tr(x_1 y))=0$. A similar argument to the one above shows that  
$$
\operatorname{tr}_{L/K}\!\big(\pi(x_1)\pi(y)\big)=0,
$$
which again implies $\pi(x_1)=0$, i.e., $x_1 \in \mathfrak{m}_S$. Thus, $x_1 = \gamma_1 x_2$ for some $x_2 \in S$, that is, $x \in \gamma_1^2 S$.  
Iterating this argument, we obtain $x \in \gamma_1^j S = \mathfrak{m}_S^j$ for every $j$. Since $\mathfrak{m}_S$ is nilpotent with nilpotency index $s_2$, it follows that $\mathfrak{m}_S^{s_2} = 0$, and hence $x = 0$, as desired.
\end{proof}
By Lemma~\ref{non-degenerate}, for any ordered basis $\underline{\beta} = (\beta_0, \beta_1, \dots, \beta_{\ell-1})$\label{dual-basis} of the free $R$-module $S$, the associated \emph{Gram matrix} defined by $$\mathrm{M}_{\underline{\beta}}=(\mathrm{T}_R^S(\beta_i, \beta_j))_{0\leq i,j< \ell}$$\label{Gram} is non-singular.
The \textit{dual basis} of $\underline{\beta}$, denoted by $\underline{\beta}^* = (\beta_0^*, \beta_1^*, \dots, \beta_{\ell-1}^*)$, is defined by $\underline{\beta}^*=\underline{\beta}\mathrm{M}_{\underline{\beta}}^{-1}$. It is easy to check that the orthogonality condition characterizes this dual basis
$$
\Tr(\beta_i\beta_j^*) = \delta_{i,j}\hbox{ for all }0 \leq i,j < \ell,
$$
where $\delta_{i,j}$ denotes the Kronecker delta. We define $R$-linear maps
\begin{align}
\mathrm{d}_{i,\underline{\beta}} : S &\to R, \quad x \mapsto \Tr(x \beta_i^*) \label{eq:d_i_alpha}, \\
\mathrm{d}_{\underline{\beta}} : S &\to R^\ell, \quad x \mapsto (\mathrm{d}_{0,\underline{\beta}}(x), \dots, \mathrm{d}_{\ell-1,\underline{\beta}}(x)). \label{eq:d_alpha}
\end{align}
Every element $x \in S$ has a unique expansion of the form $x = \sum_{j=0}^{\ell-1} c_j \beta_j$, with coefficients $c_j \in R$. Using  the $R$-linearity of the trace map and $\Tr(\beta_j \beta_k^*) = \delta_{j,k}$, we obtain
$$
\Tr(x \beta_k^*) = \Tr\left(\left(\sum_{j=0}^{\ell-1} c_j \beta_j\right) \beta_k^*\right)=\sum_{j=0}^{\ell-1} c_j \Tr(\beta_j \beta_k^*)=c_k.
$$
Therefore, $x$ can be written uniquely in the form $x = \sum\limits_{i=0}^{\ell-1} \mathrm{d}_{i,\underline{\beta}}(x) \beta_i$, which implies that $\mathrm{d}_{\underline{\beta}}$ is an $R$-module isomorphism.

\begin{example}\label{exGal} Consider the  ring $ R_\alpha = \mathbb{Z}_9[\alpha] $, where $ \alpha^2+4\alpha+8= 0 $. The polynomial $ x^2 + 4x + 8 \in \mathbb{Z}_9[x] $ is monic and basic-irreducible, so $ R_\alpha $ is a finite chain ring of length 2. In particular, $ R_\alpha $ is a Galois extension of $ \mathbb{Z}_9 $ of degree 2. The trace map $ \Tr \colon R_\alpha \longrightarrow \mathbb{Z}_9 $ is
$\Tr(x_0 + x_1 \alpha) = 2x_0 + x_1 \Tr(\alpha).$   Let $\underline{\beta} = (1, \alpha)$ be the basis of $R_\alpha$ over $\mathbb{Z}_9$, where $\alpha$ satisfies the relation $\alpha^2 = 5\alpha + 1$. The dual basis $(\alpha_0^*, \alpha_1^*)$ with respect to the trace form is obtained by $(\alpha_0^*, \alpha_1^*) = (1, \alpha) M_{\underline{\beta}}^{-1}$, where the Gram matrix $M_{\underline{\beta}}$ is given by
$$
M_{\underline{\beta}} = 
\begin{pmatrix}
\Tr(1) & \Tr(\alpha) \\
\Tr(\alpha) & \Tr(\alpha^2)
\end{pmatrix}
=
\begin{pmatrix}
2 & 5 \\
5 & 0
\end{pmatrix},
$$
with the inverse $ M_{\underline{\beta}}^{-1} = \begin{pmatrix}0 & 2 \\2 & 1\end{pmatrix}.$ Thus, the dual basis is  given $(\alpha_0^*, \alpha_1^*) = (2\alpha, 2 +  \alpha). $ This dual basis induces the map $$
\begin{array}{cccc}
\mathrm{d}_{\underline{\beta}} : & R_\alpha & \longrightarrow & (\mathbb{Z}_{9})^2 \\
  & x  & \longmapsto  & \left(2\Tr(x \alpha), 2\Tr(x)+\Tr(x \alpha)\right),
  \end{array}
$$
which is a $\mathbb{Z}_{9}$-module isomorphism. 
\end{example} 
\subsection{Eisenstein extensions of finite chain rings}\label{2}
Let $R$ be a finite chain ring with coefficient ring $\mathrm{GR}(p^e, d)$. Let $\gamma_1 \in \mathfrak{m}_R \setminus \mathfrak{m}_R^2$ denote a generator of the maximal ideal $\mathfrak{m}_R$, and $s_1$ denote the nilpotency index of $\mathfrak{m}_R$. A polynomial $g(x) \in R[x]$ of degree $k$ is called an \emph{Eisenstein polynomial} if it has the form
\begin{align}\label{eisR}
g(x) = x^k - \sum_{i=0}^{k-1} a_i x^i,
\end{align}
where each coefficient $a_i \in \mathfrak{m}_R$ for $0 \leq i < k$, and the constant term satisfies $a_0 \in \mathfrak{m}_R \setminus \mathfrak{m}_R^2$. In other words, the Eisenstein polynomial has the form 
\begin{align}\label{m23}
g(x) = x^k - \gamma_1\sum_{i=0}^{k-1} a_i x^i,  \quad a_i\in R \,\,\text{and}\,\, a_0\in R^\times.  
\end{align}
Eisenstein polynomials over \(R\) are primary and irreducible, but in general not basic-irreducible.

\begin{remark} Note that, if we want to define  a chain ring as an  Eisenstein extension of a Galois ring $\mathrm{GR}(p^e, d)$, the definition of the polynomial in Equation~\eqref{eisR} requires
 $a_i \in \langle p\rangle$ and $a_0 \in \langle p\rangle \setminus \langle p\rangle^2$.
Thus, if we factor $p$ we get the usual definition (see, for example \cite[1.5]{CL73})
$g(x) = x^k - p\sum_{i=0}^{k-1} a_i x^i,
$ where $a_i\in \mathrm{GR}(p^e, d)$ for $i=0,\ldots, k-1$, and $a_0$ is a unit.
\end{remark}

\begin{remark}\label{rem:basic} If $S|R$ is an Eisenstein extension  then $S$ and $R$ have the same residue field. Hence, given $f(x)\in R[x]$, $f(x)$ is basic irreducible over $R$ if and only if it is basic irreducible over $S$.
\end{remark}
A finite commutative ring is a chain ring if and only if it
is a homomorphic image of an Eisenstein extension of a Galois ring \cite[Theorem 2]{ClarkDrake}. More precisely in \cite{CL73}, the authors established that for every finite chain ring $R$, there exists a quintuple 
of positive integers $(p, e, d, k_1, t_1)$, which are invariants of $R$, such that
\begin{align}\label{ringR}
  R = \mathrm{GR}(p^e, d)[\gamma_1],
\end{align}
where $\mathrm{GR}(p^e,d)$ is the coefficient ring of $R$, $\gamma_1 = x + \langle g_1(x), p^{e-1} x^{t_1} \rangle$, and  $g_1(x)$\label{g_1(x)} is an Eisenstein polynomial over $\mathrm{GR}(p^e,d)$ of degree $k_1$ with $0 \leq t_1 \leq k_1$. 
The finite chain ring $R$ expressed as in  Equation~\eqref{ringR} is called  \emph{pure} if $g_1(x)=x^{k_1}-pu$, where $u$ is a unit of $\mathrm{GR}(p^e,d)$. Note that the Galois ring $\mathrm{GR}(p^e,d)$ is a finite chain ring with invariants given by the tuple $(p,e,d,1,1)$.
The extension $R | \mathrm{GR}(p^e, d)$ is  called an \emph{Eisenstein extension} of degree $k_1$.

In the same fashion, throughout this paper, we say that a ring extension $S|R$ is an \emph{Eisenstein extension of degree $k$} if there exists
an Eisenstein polynomial $g(x)$ over $R$ of degree $k$ and $0\leq t \leq k$ such that 
\begin{align}\label{ringS}
  S = R[\gamma], \quad \text{ where } \gamma = x + \langle g(x), \gamma_1^{s_1-1} x^t \rangle.
\end{align}

\begin{definition}\label{M8} Let $S | R$ be an Eisenstein extension of finite chain rings, where
   $S=R[\gamma]$\label{R[gamma]}, and $\gamma=x+\langle g(x), \gamma_1^{s_1-1}x^t\rangle$ with $g(x)$ an Eisenstein polynomial over $R$ of degree $k$, and $0\leq t\leq k$, for some $\gamma_1\in \mathfrak{m}_R\backslash\mathfrak{m}_R^2$.  We call the couple of integers  $(k,t)$ the \emph{block-degree}\label{(k,t)} of the extension $S | R$. Moreover, the extension $S | R$ is said to be \emph{pure} if  $g(x)=x^{k}-\gamma_1u$, where $u$ is a unit of $R$. 
\end{definition}

Note that there are non-isomorphic Eisenstein extensions of finite chain rings having the same degree $k$. The following example illustrates this.

\begin{example}
Consider the  ring $  \mathbb{Z}_9[\gamma] $ with $ \gamma^2 - 3 = 0 $ and $ 3\gamma = 0$. It is an Eisenstein extension of the ring $\mathbb{Z}_9$, since $\gamma=x+\langle g(x), 3x\rangle$ with $g(x)=x^2-3$.
Note that  $\mathbb{Z}_9[x]/\langle x^2 - 3, 3x\rangle$ and $\mathbb{Z}_9[x]/\langle x^2 - 6, 3x\rangle$ are two non-isomorphic Eisenstein extensions of $\mathbb{Z}_9$ of the block-degree $(2,1)$. 
\end{example}

The following proposition describes some key structural properties of Eisenstein extensions over finite chain rings.
Since the notion of independence in the context of modules and rings differs from the one known in the context of fields, we present the definition here to avoid confusion.
\begin{definition}
     Let $S | R$ be an extension of finite chain rings. The non-zero elements 
$v_1, \dots, v_k $ of $ S $ are called $R$-\emph{independent}, if for any 
$(a_1, \dots, a_k) \in R ^k,$ the equation
$
    \sum_{j=1}^k a_j v_j = 0
$
implies that $a_j v_j = 0$ for all \( j = 1, \dots, k \).
\end{definition}

\begin{prop}\label{prop2.2} Let $S | R$ be a finite chain ring extension such that $S = R[\gamma]$, where
$$
\gamma = x + \langle g(x), \gamma_1^{s_1 - 1} x^t \rangle,
$$
with $g(x)$ an Eisenstein polynomial over $R$ of degree $k$, and $\gamma_1 \in \mathfrak{m}_R \setminus \mathfrak{m}_R^2$. Suppose that $\mathfrak{m}_S$ is the maximal ideal of $S$, with nilpotency index $s_2$. Then the following statements hold.
\begin{enumerate}
  \item $\gamma\in \mathfrak{m}_S \setminus \mathfrak{m}_S^2$ and $s_2 = t + (s_1-1)k$.
   \item $\{1, \gamma, \dots, \gamma^{k-1}\}$\label{gamma-basis} is an $R$-independent generating set of $S$.
     \item  $S$ and $R$ have the same coefficient ring.
\end{enumerate} 
\end{prop} 

\begin{proof}
Let $g(x) = x^k - \gamma_1\left(\sum\limits_{i=0}^{k-1} a_i x^i\right)$ where $(a_0, \dots, a_{k-1}) \in R^k$ and $a_0 \in R^\times$.
\begin{enumerate}
    \item Since  $s_2$ is the nilpotency index of $\mathfrak{m}_S$ we have that $\gamma^{s_2-1} \neq \gamma^{s_2} = 0$ and hence $\gamma \in \mathfrak{m}_S \setminus \mathfrak{m}_S^2$.
    From the relation $g(\gamma) = \gamma_1^{s_1-1}\gamma^t = 0$, we get $\gamma^k = \gamma_1 u$ where $u \in R^\times$, and $\gamma^{t-1+(s_1-1)k} \neq\gamma^{t+(s_1-1)k} = \gamma^t \gamma_1^{s_1-1} = 0$.
    This implies that
    $$
    t + (s_1-1)k \geq s_2 > t -1 + (s_1-1)k.
    $$
    Therefore, we conclude that $s_2 = t + (s_1-1)k$.
    \item Let $a \in S$. Since $S = R[\gamma],$ we have that
$
a = \sum_{i=0}^{d} a_i \gamma^i
$
for some $(a_0, \dots, a_d) \in R^{d+1} .$
If $d \geq k$, for each $j$ with $k \leq j \leq d$, we can write
$
\gamma^j = \sum_{i=0}^{k-1} b_{i,j} \gamma^i
$
for some $b_{i,j} \in R$, and hence
$$
a = \sum_{i=0}^{k-1} \left( a_i + \sum_{j=k}^{d} a_j b_{i,j} \right) \gamma^i.
$$
This shows that $\{1, \gamma, \dots, \gamma^{k-1}\}$ is a generating set of $ S$ over $R$.  Let $(a_0, \dots, a_{k-1}) \in R^k$ such that
$
\sum_{i=0}^{k-1} a_i \gamma^i = 0.
$
Each coefficient $a_i$ can be written as $a_i = \gamma^{j_i} u_i$, where $j_i \in \{0, 1, \dots, s_2\}$ and $u_i \in R^\times$. Substituting this into the equation yields
$
\sum_{i=0}^{k-1} u_i \gamma^{i + j_i} = 0.
$
Let $ i_0 + j_{i_0} = \min\{i + j_i \mid 0 \leq i \leq k - 1\}.$ Factoring out $u_{i_0} \gamma^{i_0 + j_{i_0}}$, we obtain
\[
u_{i_0} \gamma^{i_0 + j_{i_0}} \left( 1 + \sum_{\substack{i=0 \\ i \neq i_0}}^{k-1} u_i \gamma^{i - i_0 + j_i - j_{i_0}} \right) = 0.
\]
Since $u_{i_0}$ is a unit and $\gamma$ is nilpotent, this implies $i_0 + j_{i_0} \geq s_2.$ It follows that $i + j_i \geq s_2$ for all $0 \leq i \leq k-1.$ Therefore $a_i\gamma^i=0$, for all $0 \leq i \leq k-1.$
    \item The map $$
     \begin{array}{cccc}
   \rho : & S & \rightarrow & R/\mathfrak{m}_R \\
     & \sum_{i=0}^{k-1}a_i\gamma^i  & \mapsto & x_0 +\mathfrak{m}_R
 \end{array}
    $$
is a ring epimorphism with $\mathrm{Ker}(\rho) = \mathfrak{m}_S$, since $\mathfrak{m}_R = \mathfrak{m}_S \cap R$. Consequently, $S$ and $R$ have the same characteristic and residue field. It follows that $S$ and $R$ have the same coefficient ring.
\end{enumerate}
\end{proof}

Given  an Eisenstein extension of chain rings $S|R$, to study the structure of $S$ as  an  $R$-module, we consider the quotient ring 
$\overline{R} \coloneqq R/\gamma_1^{s_1-1}R.$\label{overR}
Note that $\overline{R}$ is itself a finite chain ring whose maximal ideal $\overline{\mathfrak{m}}_R$ has nilpotency index $s_1-1$.
The natural projection
\begin{equation}
    \begin{array}{cccc}
  \,\,\overline{(\,\cdot\,)}: & R & \rightarrow & \overline{R} \\
    & x & \longmapsto & \overline{x} = x + \gamma_1^{s_1-1}R 
\end{array}
\end{equation}
\label{nat-proj}
is a ring epimorphism. Now, let us denote elements of the Cartesian product $R^t \times \overline{R}^{k-t}$\label{ambian(k,t)} as
$
\mathbf{v}=(v_0, \dots, v_{t-1} \mid \overline{v}_t, \dots, \overline{v}_{k-1}).
$
This set is an $R$-module with scalar multiplication defined by
\begin{equation}\label{Rmod}
\lambda (v_0, \dots, v_{t-1} \mid \overline{v}_t, \dots, \overline{v}_{k-1}) = (\lambda v_0, \dots, \lambda v_{t-1} \mid \overline{\lambda} \, \overline{v}_t, \dots, \overline{\lambda} \, \overline{v}_{k-1}),
\end{equation}
where $\overline{\lambda} = \lambda + \gamma_1^{s_1-1}R$ denotes the image of $\lambda$ in $\overline{R}$. 

\begin{cor} Let $0 \leq i < t \leq j < k$ and $R$ be a finite chain ring. With the notation stated above, the following maps are $R$-module isomorphisms:
\begin{align}
    \psi_i \colon \gamma^i R &\to R, & \gamma^i x &\mapsto x, \label{psi_i} \\
    \overline{\psi}_j \colon \gamma^j R &\to \overline{R}, & \gamma^j x &\mapsto x + \gamma_1^{s_1-1}R, \label{psi_j}.
\end{align}
\end{cor}

\begin{proof}
    The maps $\psi_i$ and $\overline{\psi}_j$ are $R$-module epimorphisms by construction. To prove they are isomorphisms, it suffices to show their kernels are trivial.
\begin{enumerate}
    \item   Let $\gamma^i x \in \ker(\psi_i)$ for $0 \leq i < t$. Then $0_R=\psi_i(\gamma^i x) = x $. By Proposition \ref{prop2.2}, $\gamma^i x\in S$. Thus, $\gamma^i x = \gamma^i \cdot 0_R = 0_S$, so $\ker(\psi_i) = \{0_S\}$.
    \item Let $\gamma^j x \in \ker(\overline{\psi}_j)$ for $t \leq j < k$. Then $ 0_{\overline{R}}=\overline{\psi}_j(\gamma^j x) = x+\gamma_1^{s_1-1}R $, and  $x \in \gamma_1^{s_1-1} R$. We can write $x = \gamma_1^{s_1-1} y$ for some $y \in R$. By the definition of Eisenstein polynomials in \eqref{m23}, $\gamma^k = \gamma_1 u$ with $u \in R^\times$, and hence $\gamma_1 = \gamma^k u^{-1}$. Thus
   $
       \gamma^j x = \gamma^j \gamma_1^{s_1-1} y = \gamma^j (\gamma^k u^{-1})^{s_1-1} y = \gamma^{j + k(s_1-1)} (u^{-1})^{s_1-1} y.
   $ 
   As $j \geq t$, it follows that $j + k(s_1-1) \geq t + k(s_1-1) = s_2$. Thus, $\gamma^{j + k(s_1-1)} = 0_S$, so $\gamma^j x = 0_S$. Therefore, $\ker(\overline{\psi}_j) = \{0_S\}$.
\end{enumerate}
\end{proof}

The $R$-module isomorphisms $\psi_i$ and $\overline{\psi}_j$, defined in Equations~\eqref{psi_i} and \eqref{psi_j} respectively, induce an $R$-module isomorphism given by
{\begin{align}\label{psi}
\begin{array}{r@{}c@{}ccc}
    \psi \colon & \displaystyle\prod_{i=0}^{k-1} \gamma^i R & \to & R^t \times \overline{R}^{k-t} \\[10pt]
    & (x_0, \dotsc, x_{k-1}) & \mapsto & \bigl(\psi_0(x_0), \dotsc, \psi_{t-1}(x_{t-1}) \bigm\vert \overline{\psi}_t(x_t), \dotsc, \overline{\psi}_{k-1}(x_{k-1})\bigr),
\end{array}
\end{align}}%
which is an $R$-module isomorphism. On the other hand, the fact that $\{1, \gamma, \dots, \gamma^{k-1}\}$ is an $R$-independent generating set of $S$ implies there are $R$-module epimorphisms
\begin{equation}\label{varphi-i} 
      \begin{array}{cccc}
\varphi_i : & S & \longrightarrow & \gamma^i R \\
  & x  & \longmapsto  &\varphi_i(x),
  \end{array}
\end{equation} 
for $0\leq i< k$ such that  \begin{equation}\label{varphi}
    \begin{array}{cccc}
\varphi : & S & \longrightarrow & \prod\limits_{i=0}^{k-1}\gamma^i R \\
  & x  & \longmapsto  &(\varphi_0(x),\ldots, \varphi_{k-1}(x))
  \end{array}
\end{equation} 
is an $R$-module isomorphism.  Therefore the map \begin{align}\label{xi}
                                            \xi=\psi\circ\varphi : S\longrightarrow R^t\times \overline{R}^{k-t}
                                          \end{align}
is an $R$-module isomorphism, and on the other hand,  
for all $0 \leq i < t \leq j < k$, the maps $\xi_i=\psi_i\circ\varphi_i : S\longrightarrow R$\label{xi-i} and $\overline{\xi}_j=\overline{\psi}_j\circ\varphi_j : S \longrightarrow \overline{R}$\label{xi-j}
are $R$-module epimorphisms.  

\begin{remark}
 Note that, if $t=0$ then $S=R[\gamma]\cong R^k$ as an $R$-module, where $\gamma=x+\langle g(x)\rangle$. On the other hand, if $t=k$, then $S=\overline{R}[\overline{\gamma}]\cong\overline{R}^k$ as an $\overline{R}$-module, where $\overline{\gamma}=x+\langle \overline{g}(x)\rangle$.   
\end{remark}

\begin{example}
Let $S=\mathbb{Z}_{p^a}[\gamma]$, where  $\gamma^2 -p=p^{a-1}\gamma=0$. Then $S$ is a finite chain ring. Note that the polynomial $g(x)= x^2 - p$ is Eisenstein over $\mathbb{Z}_{p^a}$, so
$S$ is an Eisenstein extension of $\mathbb{Z}_{p^a}$ with invariants $(p,a,1,2,1)$. Thus, the map $$
\begin{array}{cccc}
\xi : & S & \longrightarrow & \mathbb{Z}_{p^a}\times\mathbb{Z}_{p^{a-1}} \\
  & x_0+x_1\gamma  & \longmapsto  & (x_0, \overline{x}_1)
  \end{array}
$$
is a $\mathbb{Z}_{p^a}$-module  isomorphism. 
\end{example}
\subsection{Galois-Eisenstein extensions of finite chain rings} Throughout this subsection, we assume that $R$ is a chain ring with a quintuple  $(p, e, d, k_1, t_1)$ such that $R= \mathrm{GR}(p^e, d)[\gamma_1],$ where $\mathrm{GR}(p^e,d)$ is the coefficient ring of $R$, $\gamma_1 = x + \langle g_1(x), p^{e-1} x^{t_1} \rangle$, and  $g_1(x)\in \mathrm{GR}(p^e,d)[x]$ is an Eisenstein polynomial of degree $k_1$ with $0 \leq t_1 \leq k_1$.

Let $\gamma$ be a root of an Eisenstein polynomial $g(x) \in R[x]$ of degree $k$ such that $\gamma$ satisfies the relation $\gamma_1^{s_1 - 1} x^t$ for some $0 \leq t \leq k$. Consider the extensions $R_\alpha = R[\alpha]$ and $R_\gamma = R[\gamma]$, where
$$
    \alpha = y + \langle f(y) \rangle, \quad \text{and} \quad \gamma = x + \langle g(x),\, \gamma_1^{s_1 - 1} x^t \rangle,
$$
where $f(y) \in R[y]$ is a monic basic-irreducible polynomial  of degree $\ell$.

\begin{figure}[t]
   \begin{center}\includegraphics[width=0.75\textwidth]{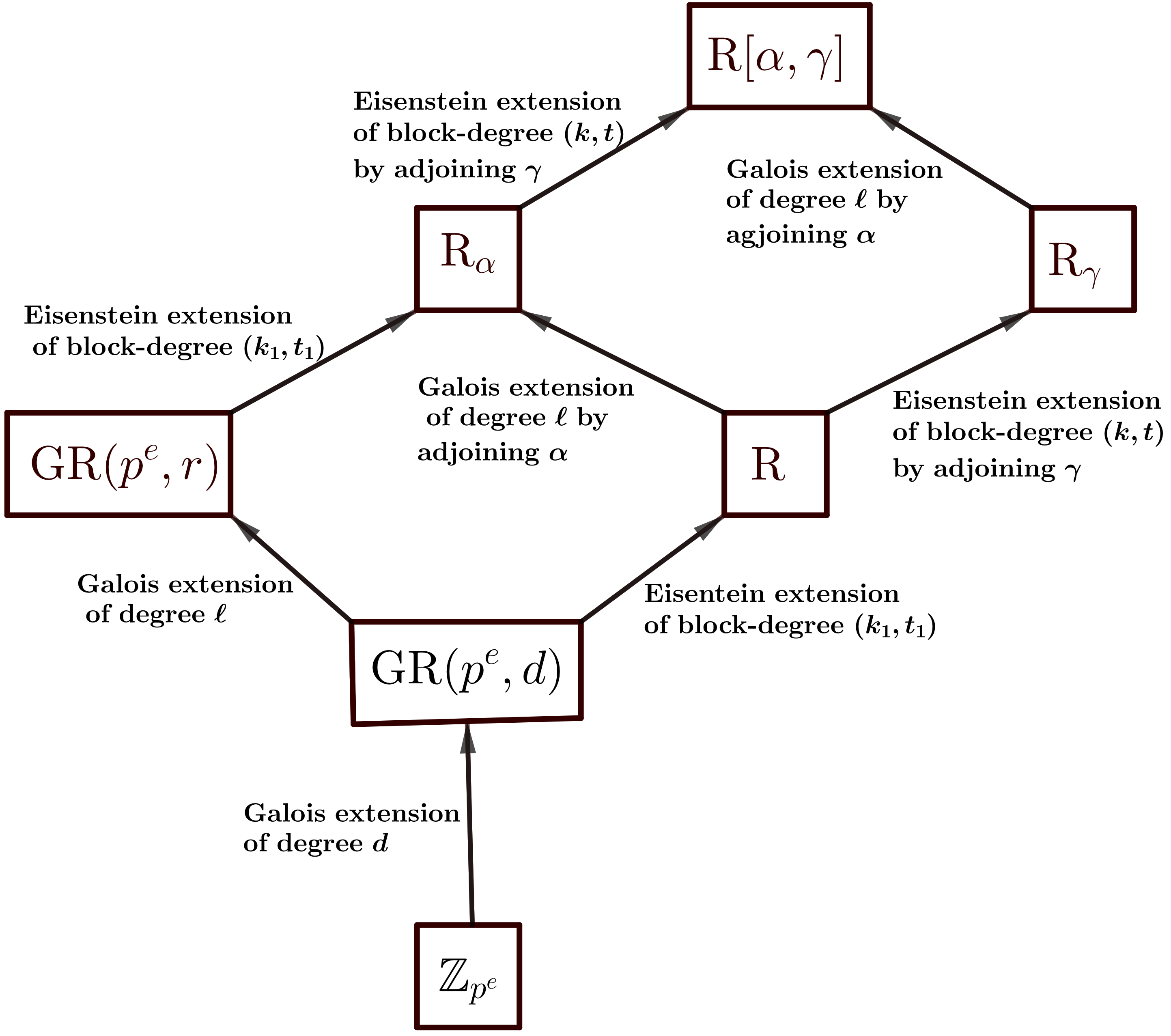} \end{center}
   \caption{Tower of extensions of a finite commutative chain ring.}\label{fig1}
\end{figure}

\begin{definition} With the notation above, an extension of finite chain rings $S|R$ is called a \emph{Galois–Eisenstein extension} \label{R[alpha,gamma]} and denoted as $S = R[\alpha, \gamma]$, if 
\begin{equation}
    R[\alpha, \gamma]=R[x,y]/\langle   g(x),\, \gamma_1^{s_1 - 1} x^t,\, f(y) \rangle.
\end{equation}
\end{definition}
Note that 
There are two ways of describing $R_\alpha[\gamma]$.
\begin{itemize}
\item \textbf{As an Einsenstein extension of a Galois ring.}  Let $R_\alpha=R[\alpha]$ be a Galois extension of $R$. By Remark \ref{m22}, $R_\alpha$ is a chain ring. Thus, there exists a quintuple  $(p, e, r, k_2, t_2)$ such that $R_\alpha = \mathrm{GR}(p^e, r)[\gamma],$ where $\mathrm{GR}(p^e,r)$ is the coefficient ring of $R_\alpha$, $\gamma = x + \langle g_2(x), p^{e-1} x^{t_2} \rangle$, and  $g_2(x)\in \mathrm{GR}(p^e,r)[x]$\label{g_2(x)} is an Eisenstein polynomial of degree $k_2$ and $0\leq t_2\leq k_2$. Since $R_\alpha$ is a Galois extension of $R$, $\mathrm{GR}(p^e,r)$ is also a Galois extension of $\mathrm{GR}(p^e,d)$ (see  Remark \ref{re-ex}), hence $d$ divides $r$. Note that  $\ell =r/d$. 
\item \textbf{As an Eisenstein extension of $R_\alpha$}. There is an Eisenstein polynomial $g_3(x)$ in  $R_\alpha[x]$ of degree  $k_3$ with $
g_3(x) = x^{k_3} - \sum_{i=0}^{k_3-1} a_i x^i,
$ 
where each coefficient $a_i \in \mathfrak{m}_{R_\alpha}$ for $0 \leq i < k_3$, and the constant term satisfies $a_0 \in \mathfrak{m}_{R_\alpha} \setminus \mathfrak{m}_{R_\alpha}^2$. Also, there exists a relation of the form $\gamma_1^{s_3 - 1} x^{t_3}$, where $0 \leq t_3 \leq k_3$, and $\langle\gamma_1\rangle=\mathfrak{m}_{R_\alpha}=\mathfrak{m}_R$ (see Remark~\ref{m22}) with nilpotency index $s_3$. 
\end{itemize}
The relationship between both of them is summarized in Figure~\ref{fig1} and the following lines.

Since $S$ is an Einsenstein extension of $\mathrm{GR}(p^e,r)$ we have that  $g_2(\gamma)=0$,  and $\gamma^k=p u$ for some $u\in \mathrm{GR}(p^e,r)^\times$. Note also that, by definition of the ring $R$,  $g_1(\gamma_1)=0$, and $\gamma_1^{k_1}=p v$ for some   $v\in \mathrm{GR}(p^e,d)^\times$.

Now, taking into account  $S$ { as an Einsenstein extension of $R_\alpha$, we have $g_3(\gamma)=0$ (since $\gamma_3$ and $\gamma$ provide the same extension $S$),  and henceforth we have that   $\gamma^{k_3}-\gamma_1 w=0$, where $w\in R_\alpha^\times$. It is also clear that $k=k_3$. Thus, $\gamma^{kk_1}=\gamma_1^{k_1}w^{k_1}=pvu^{k_1}$.   Since $\gamma^k=\gamma_1 w$, then it is a generator of the maximal ideal $\mathfrak{m}_R\subset R$, and $v u^{k_1} \in R_\alpha^\times$. The ring $R_\alpha$ is a free module over $\mathrm{GR}(p^e, d)$, thus we can write $vu^{k_1}=\sum_{i=0}^{k_1-1}a_i\gamma^{ik}$, where $a_i\in\mathrm{GR}(p^e,d)$ and $a_0\in\mathrm{GR}(p^e,d)^\times$. Hence, $\gamma$ is a root of the Eisenstein polynomial 

\begin{align}\label{m18}
  g_\alpha(x)=x^{kk_1}-p\left(\sum_{i=0}^{k_1-1}a_ix^{ik}\right)\in \mathrm{GR}(p^e,d)[x]
\end{align}
 of degree $kk_1$.   Moreover, $  \gamma^{k(s_1 - 1) + t} =\gamma^{k(s_1 - 1)} \cdot \gamma^t= \gamma_1^{s_1 - 1} u^{s_1 - 1} \cdot \gamma^t$,   and $\gamma$ is a root of $\gamma_1^{s_1 - 1} x^t$, so $\gamma$ is a root of $p^{e-1}x^{k(s_1-1)+t}$. Therefore, the nilpotency index of $\gamma_1$ is $s_1 = k_1(e-1) + t_1$. We have $k_2=kk_1$ and $t_2=k(s_1-1)+t$, and therefore 
$$S=R[\alpha,\gamma]=R_\alpha[\gamma] = \mathrm{GR}(p^e,d)[x]/\langle 
x^{kk_1}-p\left(\sum_{i=0}^{k_1-1}a_ix^{ik}\right), p x^{k(s_1-1)+t}\rangle ,$$
and then $S$ is a chain ring with coefficient ring $\mathrm{GR}(p^e,d)$.

Now, suppose that $S$ and  $R$ are chain rings and $S|R$ is an extension of finite chain rings. Then
 $$R=\mathrm{GR}(p^e,d)+\mathfrak{m}_R \mathrm{GR}(p^e,d)\hbox{ and } S=\mathrm{GR}(p^e,r)+\mathfrak{m}_S \mathrm{GR}(p^e,r).$$
 Moreover, $ S\gamma_1=S\mathfrak{m}_R\subseteq \mathfrak{m}_S$. Note that $\mathrm{GR}(p^e,d)[\alpha]=\mathrm{GR}(p^e,r)$, thus  $S=\mathrm{GR}(p^e,d)[\alpha]+\mathfrak{m}_S \mathrm{GR}(p^e,d)[\alpha]$, and $\gamma_2^k=\gamma_1 u$ for some positive integer $k$, and for some $u\in R^\times$.  Hence $\gamma_2$ is a root of an Eisenstein polynomial with coefficients in  $R$ of degree $k$, and of $\gamma_1^{s_1-1}x^{t}$ where $0\leq t\leq k$. Hence $S=R[\alpha, \gamma]$ and we say $S|R$ is a \emph{Galois-Eisenstein extension} of finite chain rings.

\begin{theorem}
\label{thm:EG} Any extension $S|R$ of finite chain rings (i.e. both $S$ and $R$ are finite chain rings) is a Galois-Eisenstein extension.  
\end{theorem}
\begin{proof} From the previous discussion, any  extension $R[\alpha,\gamma]$ of the chain ring $R$ is also  a chain ring and, any $S|R$  extension chain rings is a Galois-Einsestein $R[\alpha,\gamma]$ extension, thus, the result holds. In other words, the class of chain rings is the closure of Galois rings under Galois and Eisenstein extensions. 
\end{proof}

\begin{remark} Note that the previous result does not imply that any extension of a chain ring is a chain ring. Consider the chain ring $\mathbb{F}_2[x]/\langle x^2\rangle$ with maximal ideal and nilpotency index 2. Now, we consider its extension $\mathbb{F}_2[x,y]/\langle x^2, y^2\rangle$. We know that the ideals of the latter ring are not linearly ordered by inclusion; therefore, it is not a chain ring.
\end{remark}
The following remark clarifies the differences between the Galois extension and the Eisenstein extension.

\begin{remark}\label{rem:Rset} For an Eisenstein extension $R_\gamma|R$, we have $\mathfrak{m}_{R} R_\gamma\subsetneq\mathfrak{m}_{R_\gamma}$,  while in the Galois extension $R_\alpha|R$, we have $\mathfrak{m}_{R_\alpha}=\mathfrak{m}_{R} R_\alpha$. Note that the set 
\begin{align}\label{Rset0}
\{\alpha^i\gamma^j \;:\; 0\leq i<\ell\; \text{ and }\;  0\leq j < k \}
\end{align}
  is an $R$-independent generating set of $R[\alpha,\gamma]$.
\end{remark}
Using the isomorphisms $\mathrm{d}_{\underline{\beta}}: S\rightarrow (R_\gamma)^\ell$ and $\xi: S\rightarrow (R_\alpha)^t\times(\overline{R}_{\overline{\alpha}})^{k-t}$ defined in Equations \eqref{eq:d_alpha} and \eqref{xi}, respectively, the tower of extensions of a finite commutative chain ring in Figure~\ref{fig1} leads to the following commutative diagram
\begin{align}
\label{diag}
\begin{array}{ccc}
S & \xrightarrow{\quad \mathrm{d}_{\underline{\beta}}\quad} & (R_\gamma)^\ell \\
\Big\downarrow{\scriptstyle \xi} & \circlearrowleft & \Big\downarrow{\scriptstyle \tau \circ\xi^{\times \ell}} \\[2ex]
(R_\alpha)^t \times (\overline{R}_{\overline{\alpha}})^{k-t} & \xrightarrow{\quad \mathrm{d}_{\underline{\beta}}^{\times t} \times \overline{\mathrm{d}}_{\underline{\beta}}^{\times (k-t)} \quad} & R^{t\ell} \times \overline{R}^{(k-t)\ell},
\end{array}
\end{align}
where $\overline{\alpha}=\alpha+\gamma_1^{s_1-1}R$ and the map $\tau : (R^{t} \times \overline{R}^{(k-t)})^{\ell} \longrightarrow R^{t\ell} \times \overline{R}^{(k-t)\ell}$\label{tau} is a permutation map. Therefore, the map 
\begin{align}\label{theta}
      \theta :  S    \rightarrow    R^{t\ell} \times \overline{R}^{(k-t)\ell}
\end{align} 
defined by
$
\theta = \left(\mathrm{d}_{\underline{\beta}}^{\times t} \times \overline{\mathrm{d}}_{\underline{\beta}}^{\times (k-t)}\right) \circ \xi
$
is an $R$-module isomorphism.

 \begin{example}\label{ex_S.R} Let $\alpha = x + \langle f(x) \rangle$ and 
$\gamma = x + \langle g(x), 3 x \rangle$, where $f(x)= x^2 + 4x + 8$ is a  monic basic-irreducible over $\mathbb{Z}_{9}$, and $g(x)= x^2 - 3$ is an Eisenstein polynomial  over $\mathbb{Z}_{9}$.   The ring $\mathbb{Z}_{9}[\alpha, \gamma]$ is a Galois-Eisenstein extension of $\mathbb{Z}_{9}$ with invariants $(3,2,2,2,1)$, where  $\mathbb{Z}_{9}[\alpha]$
is a Galois extension of $\mathbb{Z}_{9}$, and  $\mathbb{Z}_{9}[\gamma]$ is an Eisenstein extension of $\mathbb{Z}_{9}$. Thus, the map   $$
\begin{array}{cccc}
\theta : & S & \longrightarrow & (\mathbb{Z}_{9})^2\times(\mathbb{Z}_{3})^2 \\
  & x_0+x_1\gamma  & \longmapsto  & \left(  \mathrm{d}_{\underline{\beta}}(x_0) \,|\, \overline{\mathrm{d}}_{\underline{\beta}}(\overline{x}_1)\right),
  \end{array}
$$ 
is a $\mathbb{Z}_{9}$-module  isomorphism. 
\end{example}
\section{Group rings}\label{sec:gR}
In this section, $S|R$ is an extension of finite chain rings, $C$ is a non-empty subset of $S^n$, and $G$\label{G} is a finite group (written multiplicatively) of order $n$.  According to Cayley's Theorem,  $G$ can be embedded into a subgroup of the symmetric group $\mathbb{S}_n$.\label{Sn} 
\subsection{General properties of group rings}\label{subsec:gR}

The \emph{group ring} of $G$ over $S$,  denoted $S[G]$, is the set of all formal sums
$$
S[G] = \left\{ \sum_{g\in G} a_{g} g \mid a_g\in S \right\}, $$\label{SG}
with componentwise addition:
$$
\left( \sum_{g \in G} a_g g \right) + \left( \sum_{g \in G} b_g g \right) = \sum_{g \in G} (a_g + b_g) g,
$$
and multiplication is defined by bilinearly extending the rule
$$
(a_g g)(b_h h) = (a_g b_h)(gh),
$$
so that
$$
\left( \sum_{g \in G} a_g g \right) \left( \sum_{h \in G} b_h h \right) = \sum_{k \in G} \left( \sum_{\substack{g, h \in G \\ gh = k}} a_g b_h \right) k.
$$
$S[G]$ is a ring with identity element $1 = 1_S 1_G$,\label{1} where $1_G$ is the identity element of $G$, and with the zero element  $0 = \sum_{g \in G} 0_S g$. 

From now on in the paper, we will denote the group rings $S[G]$ and $R[G]$\label{RG} by $\mathscr{S}$\label{notaSG} and $\mathscr{R}$,\label{notaRG} respectively. When convenient, we will consider an ordering on the elements of $G$ as follows: $G = \{g_1, \dots, g_n\}$, where $g_1=1_G$ is the identity element in $G$. Accordingly, we will use two different notations for elements in the group ring:  $\sum_{g\in G} a_g g$ and $\sum_{i=1}^{n} a_{g_i} g_i$, depending on whether we want to emphasize the order given on the group elements. 
 
Note that the group $G$ can be viewed as a subgroup of the group of units of $\mathscr{S}$, and the map    
\begin{align}\label{eq:mu} 
\begin{array}{cccc}
\mu : & \mathscr{S}                & \longrightarrow & \mathscr{S} \\
          &\sum\limits_{g\in G} a_{g} g & \longmapsto & \sum\limits_{g\in G}a_{g} g^{-1} 
\end{array} 
\end{align} 
is an involutive ring anti-automorphism that fixes $S$ pointwise. Hence, the image of a left ideal in $\mathscr{S}$  under $\mu$ is a right ideal in $\mathscr{S}$, and vice versa.
In addition to its ring structure, $\mathscr{S}$ becomes an $S$-bimodule when equipped with the left and right scalar multiplications given by
{\begin{align*}
s \left( \sum_{g \in G} a_g g \right) = \sum_{g \in G} (s a_g) g  \;\;\text{and }
\left( \sum_{g \in G} a_g g \right) s = \sum_{g \in G} (a_g s) g,
\end{align*}}%
for all $s\in S$. Thus, $\mathscr{S}$ is a free $S$-bimodule with basis $G$.  For further details on group rings, we refer the reader to \cite{MS92, Pas77}.  

Finally, the group ring $\mathscr{S}$ becomes an $\mathscr{R}$-bimodule when endowed with the natural left and right scalar multiplications by elements of $\mathscr{R}$.

Note that the image under $\mu$ of any left $\mathscr{R}$-submodule of $\mathscr{S}$ is a right $\mathscr{R}$-submodule of $\mathscr{S}$, and vice versa. 
The natural projection $\pi \colon R \to \mathbb{F}_{p^d}$ extends componentwise to the group ring $\mathscr{S} = S[G]$ as follows,
\begin{align}\label{pi_G}
  \begin{array}{cccc}
  \pi: & \mathscr{S} & \longrightarrow  & \mathbb{F}_{p^r}[G]\\
    & \sum\limits_{g \in G} a_g g & \longmapsto & \sum\limits_{g \in G} \pi(a_g) g,
\end{array}
\end{align} 
where $\pi$ on the right-hand side is the restriction of the projection $S \to \mathbb{F}_{p^r}$. This extension is a ring epimorphism.

The ring epimorphism
\begin{align}\label{aug}
  \begin{array}{cccc}
  \text{Aug}_\mathscr{S} : & \mathscr{S}                 & \longrightarrow & S\\
                           & \sum\limits_{g \in G} a_g g & \longmapsto     & \sum\limits_{g \in G} a_g
\end{array}
\end{align} 
is called \emph{augmentation map} with respect to $\mathscr{S}$, and its kernel \begin{align}\label{ideal_aug}
    \mathfrak{a}_\mathscr{S} = \left\{ \sum\limits_{g \in G} a_g g \in \mathscr{S} \mid \sum\limits_{g \in G} a_g = 0 \right\}
    \end{align} is a two-sided ideal in $\mathscr{S}$ and is called the \emph{augmentation ideal} of $\mathscr{S}$.  Consider the projection map
\begin{align}\label{coef}
  \begin{array}{cccc}
  \coeffId{\,\cdot\,}: & \mathscr{S} & \longrightarrow  & S\\
    & \sum\limits_{g \in G} a_g g & \longmapsto & a_{1_G}, 
\end{array}
\end{align} 
which is an $S$-module epimorphism. It is clear that  for $a = \sum_{g \in G} a_g g$, we have $\coeffId{g^{-1}a} = a_g$. Therefore
\begin{align}\label{Eq_1G}
    a= \sum\limits_{g \in G} \coeffId{g^{-1}a} g.
\end{align} 
 In addition, the free $S$-module $\mathscr{S}$ carries a symmetric, non-degenerate, and $G$-invariant bilinear form $\langle \cdot, \cdot \rangle$ which is determined by
\begin{align}\label{inner}
\langle g, h \rangle = \delta_{g,h}=
\begin{cases} 
1 & \text{if } g = h \\
0 & \text{otherwise},
\end{cases}
\end{align}
for all $g, h \in G$. Here, $G$-invariance means that $\langle ag, bg \rangle = \langle a, b \rangle$ for all $(a, b) \in \mathscr{S}^2$ and all $g \in G$. 
It is easy to see that, for any pair $(a, b) {=(\sum_{g\in G} a_g g, \sum_{g\in G} b_g g)} \in \mathscr{S}^2$, the following relation holds:
\begin{align}\label{eq:dot_product}
    \langle a, b\rangle = \coeffId{a\mu(b)} = \coeffId{\mu(a)b} { {=\coeffId{\sum_{g,h\in G} a_g b_h\, g h^{-1}}=\sum_{g\in G} a_g b_g}},
\end{align}
{since $gh^{-1}=1_G$ implies $h=g$.}
Via the $S$-module isomorphism  
\begin{align}\label{PhiO}
 \begin{array}{cccc}
  \Phi: & S^n & \longrightarrow & \mathscr{S} \\
          & (a_{g_1}, \dots, a_{g_n}) & \longmapsto & \sum\limits_{i=1}^{n} a_{g_i} g_i,
\end{array}
\end{align}
the bilinear form defined in Equation \eqref{inner} corresponds to the usual Euclidean inner product on $S^n$. 

Recall that the $R$-module $S^n$ is an $R$-bimodule  equipped with the left and right scalar multiplications
$$\lambda(x_1,\ldots, x_n)=(x_1,\ldots, x_n)\lambda,$$
since $\lambda x_i=x_i\lambda$, for all $1\leq i \leq n$. Note that every $ R$-submodule of $S^n$ naturally inherits an $ R$-bimodule structure. These particular $R$-submodules of $S^n$ are called  {$S|R$-\emph{additive codes} of length $n$.}

\begin{definition} An {$S|R$-\emph{additive code}\label{defi-add} of length $n$} is a
\begin{enumerate}
  \item \emph{left additive $G$-code} (left additive group code with respect to $G$) if it is invariant under the left action of $G$ on $S^n$, given  by 
\begin{align}\label{def:left_action}
^h(a_{g_1}, \dots, a_{g_n})   = (a_{hg_1}, \dots, a_{hg_n})
\end{align}
for all $h \in G$ and all $(a_{g_1}, \dots, a_{g_n}) \in S^n$.
  \item \emph{right additive $G$-code} (right additive group code with respect to $G$) if it is invariant under the right action of $G$ on $S^n$, defined by 
\begin{align}\label{def:right_action}
(a_{g_1}, \dots, a_{g_n})^h   = (a_{g_1h}, \dots, a_{g_nh})
\end{align}
for all $h \in G$ and all $(a_{g_1}, \dots, a_{g_n}) \in S^n$.
\item \emph{two-sided additive $G$-code} (two-sided additive group code with respect to $G$) if it is a left and a right additive $G$-code.
\end{enumerate}
\end{definition}

If $C$ is an (left or right) additive $G$-code, then the group $G$ is a subgroup of  
$$
\mathrm{PAut}(C) = \{\tau \in \mathbb{S}_n \mid \tau(C) = C\},
$$\label{paut}%
where
$ 
\tau(C) = \left\{(c_{g_{\tau(1)}}, \dots, {g_{\tau(n)}}) \mid (c_{g_1}, \dots, c_{g_n}) \in C\right\}.
$ 
Note that $\Phi(a^g) = \Phi(a) g^{-1}$  holds for all $g \in G$ and all $a = \sum_{i=1}^n a_{g_i} g_i \in \mathscr{S}$ since
\begin{align}\label{eq:right_action}
\Phi(a^g) = \sum_{i=1}^n a_{g_i g} g_i = \sum_{j=1}^n a_{g_j} (g_j g^{-1}) = \left( \sum_{j=1}^n a_{g_j} g_j \right) g^{-1} = \Phi(a) g^{-1},
\end{align}
because $\Phi: S^n \rightarrow \mathscr{S}$ is an $S$-module isomorphism. Similarly, we have the corresponding identity for the left action
\begin{align}\label{eq:left_action}
    \Phi(^g a) = g^{-1} \Phi(a),
\end{align}
which holds for all $g \in G$ and all $a \in \mathscr{S}$.

The following key proposition establishes a correspondence between left $S| R$-additive $G$-codes and left $\mathscr{R}$-submodules of $\mathscr{S}$. 

\begin{prop}\label{prop:main}
Let $C$ be an $S|R$-additive code. Then $C$ is a left $G$-code if and only if $\Phi(C)$ is a left $\mathscr{R}$-submodule of $\mathscr{S}$.
\end{prop}

\begin{proof}
The map $\Phi$ is an $R$-module isomorphism since $R$ is a subring of $S$. Thus $C$ is a left code if and only if $\Phi(C)$ is a left $R$-submodule of $\mathscr{S}$. Moreover, the equality $\Phi(^g a) = g^{-1} \Phi(a),$ defined in Equation~\eqref{eq:left_action} for all $g \in G$ and all $a \in \mathscr{S}$, is equivalent to  $\Phi(C)$ being a left $\mathscr{R}$-submodule of $\mathscr{S}$. 
\end{proof} 

\begin{example}\label{ExG-code1}
Let $S=R[\alpha]$ be a finite chain ring with $R=\mathbb{Z}_9$, the element $\alpha$  given as in Example~\ref{ex_S.R}, and $\mathbb{S}_3 = \{1, \varrho_1, {\varrho_1}^2, {\varrho_2}, \varrho_1, \varrho_2, \varrho_1^2\varrho_2\}$ be the symmetric group  where $\varrho_1 = (123)$ and $\varrho_2 = (12)$.  
\begin{enumerate}
    \item Consider the set  $C_1 = \Phi^{-1}(\alpha\mathfrak{a}_\mathscr{R})$ where $\mathfrak{a}_\mathscr{R}$ is defined in Equation~\eqref{ideal_aug}. Since $\mathfrak{a}_\mathscr{R}$ is two-sided ideal,  $C_1$ is a  two-sided additive $\mathbb{S}_3$-code.
    \item Consider the set $C_2 \subseteq S^6$ such that $\Phi(C_2) = \mathscr{R}\alpha(1-\varrho_2)$. This is a left ideal in $\mathscr{S}$, so by Proposition \ref{prop:main}, $C_2$ is a left additive $\mathbb{S}_3$-code. However, $C_2$ is not a right additive $\mathbb{S}_3$-code. To show this, choose an element $a=\alpha(1-\varrho_2)\in \mathscr{R}\alpha(1-\varrho_2)=\Phi(C_2)$. Assume that  $a\varrho_2^2\in \mathscr{R}\alpha(1-\varrho_2)$. Then there exists an element $y\in \mathscr{R}$ such that $(1-\varrho_2)\varrho_1^2=y (1-\varrho_2)$.  List the group elements of $\mathbb{S}_3$ as $g_1 = 1,\, g_2 = \varrho_1,\,  g_3 = \varrho_1^2,\, g_4 = \varrho_2,\, g_5 = \varrho_1 \varrho_2,\, g_6 = \varrho_1^2 \varrho_2.$ Then $ (1 - \varrho_2) \varrho_1^2 = \varrho_1^2 - \varrho_1 \varrho_2=g_3 - g_5.$  Assume that  $y = \sum_{g \in \mathbb{S}_3} y_g g$. Thus 
    \begin{align*}
       y - y \varrho_2 = (c_1 - c_4)g_1 &+ (c_2 - c_5)g_2 + (c_3 - c_6)g_3\\ 
        &+ (c_4 - c_1)g_4 + (c_5 - c_2)g_5 + (c_6 - c_3)g_6.
    \end{align*}
Equating the coefficients of the both sides of $(1 - \varrho_2) \varrho_1^2 = y (1 - \varrho_2)$, we obtain  $c_5 - c_2=0$ and  $c_5 - c_2 = -1$, which is a contradiction. Therefore, $a\varrho_2^2\notin \mathscr{R}\alpha(1-\varrho_2)$. This means we found an element $a \in \Phi(C_2)$ and a group element $\varrho_2^2 \in \mathbb{S}_3$ such that $a\varrho_2^2\notin \Phi(C_2)$, which proves that $\Phi(C_2)$ is not a right $\mathscr{R}$-submodule of $\mathscr{S}$. Hence, by Proposition \ref{prop:main}, $C_2$ is not a right additive $\mathbb{S}_3$-code.
 \end{enumerate}   
\end{example}

Note that if $C$ is a left $S|R$-additive $G$-code, then $\pi(C)$ is also a left $\mathbb{F}_{p^r}|\mathbb{F}_{p^d}$-additive $G$-code.  
For the rest of this paper, we identify left $S|R$-additive $G$-codes with left $\mathscr{R}$-submodules of $\mathscr{S}$ via the isomorphism $\Phi$. Since the image of a right \( S|R \)-additive \( G \)-code under the anti-automorphism \( \mu: \mathscr{S} \rightarrow \mathscr{S} \) is a left \( S|R \)-additive \( G \)-code, in what follows we focus our attention on the latter ones.

\subsection{CRT decompositions of the Abelian group ring}\label{ssec:CRT}
When the group $G$ is commutative, one can decompose the group algebra by considering the cyclotomic classes associated with each cyclic component, using the Chinese Remainder Theorem and the Fourier transform. See, for example, \cite{Fourier} and the references therein. This decomposition was also used in \cite{MOO18} for dealing with the additive cyclic case. We will extend these techniques to the Abelian group algebras in the serial module case, i.e., $\gcd(|G|, p) = 1$. Note that semisimplicity in the finite field case becomes seriality for finite chain rings (see \cite{Serial} or \cite{Puninski} for a more comprehensive account of this fact).

Throughout this  subsection, $S|R$ will be  an extension of finite chain rings of parameters $(p,s_1, \ell, k, t)$, and $G$ a finite abelian group of order $n$, with $\gcd(n, p) = 1$, written as a direct product of $\flat\in\mathbb N$ cyclic groups  $G  = \langle c_{_{1}} \rangle \times
\cdots \times \langle c_{_{\flat}} \rangle,$ where $ c_a\in G,\, a=1,\ldots, \flat$\label{decom_G}. If we denote by $n_a$\label{n_a}  the order  of the cyclic group $\langle c_{_a} \rangle$, then $n=|G|=\prod_{a=1}^\flat n_a$. 
It is well-known that the group rings  $\mathscr{R}$ and $\mathscr{S}$ decompose as \begin{equation}
 \mathscr{R}=\bigotimes\limits_{a=1}^{\flat}\mathscr{R}_a\text{ and }\mathscr{S}=\bigotimes\limits_{a=1}^{\flat}\mathscr{S}_a,   
\end{equation}
where $\mathscr{R}_a=R[\langle c_{_a} \rangle]$\label{R_a} and $\mathscr{S}_a=S[\langle c_{_a} \rangle]$\label{S_a} for  $a=1,\ldots, \flat$. 

\begin{definition}
Let $1 \leq a \leq \flat $, $b \in \{q, q^\ell\} $, and $0 \leq j < \ell $. 
The set
$$
    \complement_{i,j,a}^{(b)} = \left\{ i q^j b^h \bmod n_a \mid h \in \mathbb{Z}^+ \right\}
$$\label{cyclo}
is called the \emph{$b$-cyclotomic coset} modulo $n_a$.
\end{definition}
The size of the \emph{$b$-cyclotomic coset} modulo $n_a$ is denoted by $\kappa_{i,j,a}^{(b)} = \lvert \complement_{i,j,a}^{(b)} \rvert $\label{Order_cyclo}. The following properties hold:
\begin{enumerate}
    \item If $\gcd(\kappa_{i,0,a}^{(q)}, \ell) = 1$, then $\complement_{i,j,a}^{(q^\ell)} = \complement_{i,0,a}^{(q)}$.
    \item If $\ell$ divides $\kappa_{i,0,a}^{(q)}$, then $\kappa_{i,0,a}^{(q)} = \ell\kappa_{i,j,a}^{(q^\ell)}$ and $\complement_{i,0,a}^{(q)} = \bigcup\limits_{j=0}^{\ell-1}\complement_{i,j,a}^{(q^\ell)}$.
\end{enumerate}
Let $Z_a^{(q)}=\{i_0,\ldots,i_{v_a}\}$\label{Z_a^q} be a complete set of representatives of all the distinct $q$-cyclotomic cosets modulo $n_a$, where $0=i_0< i_1 < \cdots < i_{v_a} < n_a$, such that $\gcd(\kappa_{i_x,0,a}^{(q)}, \ell) = 1$ for $0 \leq x \leq u_a$, and $\kappa_{i_x,0,a}^{(q)}$ is a multiple of $\ell$ for $u_a + 1 \leq x \leq v_a$.
Then $\complement_{i_0,0,a}^{(q)},\, \complement_{i_1,0,a}^{(q)},\, \ldots, $ and $ \complement_{i_{v_a},0,a}^{(q)}$ are all distinct $q$-cyclotomic cosets modulo $n_a$ and
\[ \mathbb{Z}_{n_a} = \bigsqcup_{x=0}^{v_a} \complement_{i_x,0,a}^{(q)} \]
is a disjoint union. Hence, $\complement_{i_x,0,a}^{(q)} = \complement_{i_x,0,a}^{(q^\ell)}$ for $0 \leq x \leq u_a$, $\complement_{i_x,0,a}^{(q)} \neq \complement_{i_x,0,a}^{(q^\ell)}$ and $$\complement_{i_x,0,a}^{(q)} = \complement_{i_x,0,a}^{(q^\ell)} \cup \complement_{i_x,1,a}^{(q^\ell)} \cup \cdots \cup \complement_{i_x,\ell-1,a}^{(q^\ell)}$$ for $u_a + 1 \leq x \leq v_a$. 
  Moreover, by suitably choosing representatives of the $q$-cyclotomic cosets modulo $n_a$, we can assume, without any loss of generality, that
$Z_{a}^{(q)} = F_{1,a}^{(q)} \sqcup F_{2,a}^{(q)}\sqcup L_{1,a}^{(q)} \sqcup L_{2,a}^{(q)} $
(a disjoint union), where 
\begin{align*}
F_{1,a}^{(q)} &= \left\{i \in Z_a^{(q)} \mid i_0\leq i \leq i_{u_a} \text{ and } \complement_{i,0,a}^{(q)} = \complement_{-i,0,a}^{(q)}\right\}, \\
F_{2,a}^{(q)} &= \left\{i \in Z_a^{(q)} \mid i_{u_a}< i \leq i_{v_a} \text{ and } \complement_{i,0,a}^{(q)} = \complement_{-i,0,a}^{(q)}\right\}, \\
L_{1,a}^{(q)}  &= \left\{i\in Z_a^{(q)} \mid i_0\leq i \leq i_{u_a} \text{ and } \complement_{i,0,a}^{(q)} \neq \complement_{-i,0,a}^{(q)}\right\},\\
L_{2,a}^{(q)}  &= \left\{i\in Z_a^{(q)} \mid i_{u_a}< i \leq i_{v_a} \text{ and } \complement_{i,0,a}^{(q)} \neq \complement_{-i,0,a}^{(q)}\right\}.
\end{align*}

\begin{example} Consider $q = 3$, $\ell = 2$ and $G=\mathbb Z_8\times \mathbb Z_4$, thus $\flat=2$. 
\begin{enumerate}
    \item  $n_1 = 8 $  \begin{itemize}
  \item For $0 \leq x \leq u_2 = 1$, we have:\\ $i_0 = 0$: $\complement_{0,0,1}^{(3)} = \{0\} = \complement_{0,0,1}^{(9)}$,\\
   $i_1 = 4$: $\complement_{4,0,1}^{(3)} = \{4\} = \complement_{4,0,1}^{(9)}$.
  \item For $u_2 + 1 = 2 \leq x \leq v_2 = 4$, we have:\\ $i_2 = 1$: $\complement_{1,0,1}^{(3)} = \{1, 3\} = \complement_{1,0,1}^{(9)} \cup \complement_{1,1,1}^{(9)} = \{1\} \cup \{3\}$,\\ $i_3 = 2$: $\complement_{2,0,1}^{(3)} = \{2, 6\} = \complement_{2,0,1}^{(9)} \cup \complement_{2,1,1}^{(9)} = \{2\} \cup \{6\}$,\\ $i_4 =5$: $\complement_{5,0,1}^{(3)} = \{5, 7\} = \complement_{5,0,1}^{(9)} \cup \complement_{5,1,1}^{(9)} = \{5\} \cup \{7\}$.
\end{itemize}
Thus $Z_{1}^{(3)}=\{0, 1, 2, 4, 5\}$, { $F_{1,2}^{(3)}  = \{0, 4\}$, $F_{2,2}^{(3)}= \{2\}$, $L_{1,2}^{(3)} = \{1, 5\}$.}
  \item $n_2 = 4 $ \begin{itemize}
  \item For $0 \leq x \leq u_3 = 1$, we have:\\ $i_0 = 0$: $\complement_{0,0,2}^{(3)} = \{0\} = \complement_{0,0,2}^{(9)}$,\\
   $i_1 = 2$: $\complement_{2,0,3}^{(3)} = \{2\} = \complement_{2,0,2}^{(9)}$.
  \item For $u_3 + 1 = 2 \leq x \leq v_3 = 2$, we have:\\ $i_2= 1$: $\complement_{1,0,2}^{(3)} = \{1, 3\} = \complement_{1,0,2}^{(9)} \cup \complement_{1,1,2}^{(9)} = \{1\} \cup \{3\}$.
\end{itemize}
Thus $Z_{2}^{(3)}=\{0, 1, 2\}$, $F_{2,2}^{(3)} = \{0, 2\}$ and $L_{2,2}^{(3)} =\{1\}$.
\end{enumerate}
 
\end{example}
Idempotent elements of rings (under multiplication) play a central role in the decomposition of modules, particularly when the idempotents are primitive, meaning that the corresponding summands are indecomposable. An idempotent is said to be central if it commutes with every element of the ring. For precise definitions and their applications to linear codes, see \cite{MS92} and the references therein. In what follows, we explicitly construct idempotent elements in the setting of serial Abelian group rings and use them to describe $S|R$-additive codes over these rings.
 
Let $\widehat{S}_a$ be a Galois extension of $S$ of degree $m_a$, where  $m_a= \min\{j \in \mathbb{Z}^+ \mid (q^\ell)^j \cong 1 \bmod n_a\}$. We  denote by $\zeta_a$ a generator of the cyclic group $\Gamma(\widehat{S}_a)^*$, and by $\eta_a= \zeta_a^{\frac{q^{\ell m_a}-1}{n_a}}$ a primitive $n_a$th root of unity.  The \emph{primitive idempotent elements} of the ring $\mathscr{R}_a$ are given by
\begin{align}\label{idem_R}
\epsilon_{i,a} = \frac{1}{n_a} \sum_{y=0}^{n_a-1} \left(\sum_{z \in \complement_{i,0,a}^{(q)} } \eta_a^{-yz}\right) c_{a}^{y}
\end{align}
for $ i \in Z_a^{(q)} $. We  denote by $\mathscr{K}_{i,a} = \epsilon_{i,a}\mathscr{R}_a$, for each $ i \in Z_a^{(q)}$. It is easy to check that $\mathscr{K}_{i,a}$ is a finite chain ring with invariants $(p, e_1, d \kappa_{i,0,a}^{(q)}, k_1, t_1)$ with identity $\epsilon_{i,a}$. Moreover, 
$$
\mu(\epsilon_{i,a})=\left\{
                        \begin{array}{ll}
                          \epsilon_{i,a}, & \hbox{if $i\in F_{2,a}^{(q)}$,} \\
                          \epsilon_{-i,a}, & \hbox{if $i\in L_{2,a}^{(q)}$.}
                        \end{array}
                      \right.
$$

\begin{lemma}[Lemma 4.1 in \cite{MOO18}]\label{lem:idempotent-Sa} The set of all pairwise orthogonal primitive central idempotent elements of $\mathscr{R}_a$ is given by $\{\epsilon_{i,a} \mid  i\in Z_a^{(q)}\},$ and they provide the following decomposition  $$\mathscr{R}_a =\left(\bigoplus\limits_{i\in F_{1,a}^{(q)}}\mathscr{K}_{i,a}\right)\oplus\left(\bigoplus\limits_{i\in L_{1,a}^{(q)}}\mathscr{K}_{i,a}\right)\oplus\left(\bigoplus\limits_{i\in F_{2,a}^{(q)}}\mathscr{K}_{i,a}\right)\oplus\left(\bigoplus\limits_{i\in L_{2,a}^{(q)}}\mathscr{K}_{i,a}\right).$$
\end{lemma}

\begin{cor}\label{thm:idempotent_R}   Let  
  $\mathbf{x}=({x_1},\ldots, {x_\flat})\in Z^{(q)}=\prod\limits_{a=1}^\flat Z_{a}^{(q)}$ and $
\epsilon_{\mathbf{x}} = \epsilon_{{x_1}, 1} \cdots \epsilon_{{x_\flat}, \flat}
$.
Then $$\epsilon_{{\mathbf{x}}} =\frac{1}{n}\sum_{y_1=1}^{n_1} \cdots \sum_{y_\flat=1}^{n_\flat} 
\left( 
    \prod_{a=1}^{\flat} 
    \left( 
        \sum_{z_a \in \complement_{{x_a},0,a}^{(q)}} 
        \eta_a^{-y_a z_a} 
    \right) 
\right) 
c_1^{y_1} \cdots c_\flat^{y_\flat}$$
and
$
\{\epsilon_{{\mathbf{x}}} \mid \mathbf{x}\in Z^{(q)} \}
$
is the set of pairwise orthogonal primitive central idempotent elements of $\mathscr{R}$, and  $\mathscr{R} =\bigoplus\limits_{\mathbf{x}\in Z^{(q)}}  \mathscr{K}_{\mathbf{x}} $, where $\mathscr{K}_{\mathbf{x}}=\mathscr{R}\epsilon_{{\mathbf{x}}}$.
\end{cor}

\begin{example} Consider the abelian group $G=\langle c_{_1} \rangle\times \langle c_{_2} \rangle$ where $\langle c_{_1} \rangle$ and $\langle c_{_2} \rangle$ are the multiplicative cyclic groups of order $8$ and $4$, respectively. Let $S$ be an extension of $\mathbb{Z}_9$ with invariants $(3,2,2,2,1)$.  Then $q=3, \ell=2, m_2=m_3=1$ and $\Gamma(\widehat{S}_2)^*=\Gamma(\widehat{S}_3)^*=\langle\zeta\rangle$, where $\zeta$ is a root of the basic primitive polynomial $x^2 + 4x + 8$ over $\mathbb{Z}_9$. Thus $\eta_2=\zeta$ and $\eta_3=\zeta^2$. Therefore, the pairwise orthogonal primitive central idempotent elements of $\mathbb{Z}_9[\langle c_{_1} \rangle]$ are
\begin{eqnarray*} 
\epsilon_{0,1} &=& -(1+c_{_1}+c_{_1}^2+c_{_1}^3+c_{_1}^4+c_{_1}^5+c_{_1}^6+c_{_1}^7),\\ 
 \epsilon_{4,1} &=& 1-c_{_1}+c_{_1}^2-c_{_1}^3+c_{_1}^4-c_{_1}^5+c_{_1}^6-c_{_1}^7,\\ 
 \epsilon_{1,1} &=& 7+5c_{_1}+5c_{_1}^3+7c_{_1}^4+4c_{_1}^5+4c_{_1}^7,\\  
 \epsilon_{2,1} &=& 7+2c_{_1}^2+7c_{_1}^4+2c_{_1}^6,\\
\epsilon_{5,1} &=& 7+4c_{_1}+4c_{_1}^3+2c_{_1}^4+5c_{_1}^5+5c_{_1}^7,
\end{eqnarray*} 
and the pairwise orthogonal primitive central idempotent elements of $\mathbb{Z}_9[\langle c_{_2} \rangle]$ are
\begin{eqnarray*} 
\epsilon_{0,2} &=& 7(1+c_{_2}+c_{_2}^2+c_{_2}^3),\\
 \epsilon_{2,2} &=& 7(1-c_{_2}+c_{_2}^2-c_{_2}^3),\\
 \epsilon_{1,2} &=& 5(1-c_{_2}^2).\\
 \end{eqnarray*} Hence, the pairwise orthogonal primitive central idempotent elements of $\mathbb{Z}_9[G]$ are given by $\epsilon_{{i_1}, 1}\epsilon_{{i_2}, 2}$ for all $(i_1,i_2)\in \{0, 1, 2, 4, 5\}\times\{0, 1, 2\}$. 
\end{example} 

The idempotent elements of the group ring $\mathscr{S}_a$ are given by
\begin{align}\label{idem_S}
\epsilon_{i,j,a} = \frac{1}{n_a} \sum_{y=0}^{n_a-1} \left(\sum_{z\in \complement_{x,j,a}^{(q^\ell)} } \eta_a^{-yz}\right) c_{a}^y,  \hbox{ for } i \in F_{2,a}\cup L_{2,a},\, 0 \leq j <\ell, 
\end{align}
where  $a=1,\ldots , \flat$.

\begin{example} Consider the multiplicative cyclic group $\langle c_{_1} \rangle$ of order $8$, and let  $S$ be an 
extension of $\mathbb{Z}_9$ with invariants $(3,2,2,2,1)$. Then $q=3, \ell=2, m_2=1$ and $\Gamma(S)^*=\langle\zeta\rangle$, where $\zeta$ is a root of the basic primitive polynomial $x^2 + 4x + 8$ over $\mathbb{Z}_9$. Thus $\eta_2=\zeta$ and the primitive idempotent elements of $S[\langle c_{_1} \rangle]$ are:
\[
\begin{aligned}
\epsilon_{1,0,1}  =  \sum_{y=0}^{7} 8\zeta^{-y} c_{_1}^y,&  &\epsilon_{1,1,1}  =  \sum_{y=0}^{7} 8\zeta^{-3y} c_{_1}^y,& &\epsilon_{2,0,1}  =  \sum_{y=0}^{7} 8\zeta^{-2y} c_{_1}^y, \\
\epsilon_{2,1,1}  = \sum_{y=0}^{7} 8\zeta^{-6y} c_{_1}^y,&  & \epsilon_{5,0,1} = \sum_{y=0}^{7} 8\zeta^{-5y} c_{_1}^y,& & \epsilon_{5,1,1}  =  \sum_{y=0}^{7} 8\zeta^{-7y} c_{_1}^y.
\end{aligned}
\]
\end{example}

If we denote by $\mathscr{L}_{i,a} = \epsilon_{i,a}\mathscr{S}_a$ for all $i \in Z_{a}^{(q)}$, we have that 
 $\mathscr{K}_{i,a} = \mathscr{L}_{i,a} \cap \mathscr{R}_a$ for all $ i \in Z_{a}^{(q)}$, and $\mathscr{L}_{i,a}$ is a finite principal ideal ring with identity $\epsilon_{i,a}$. On the other hand, let $\mathscr{L}_{i,j,a} = \epsilon_{i,j,a}\mathscr{S}_a$ and  $\mathscr{K}_{i,j,a} = \epsilon_{i,j,a}\mathscr{R}_a$, for all $i \in F_{2,a}^{(q)}\cup L_{2,a}^{(q)}$ and $0 \leq j <\ell$. By \cite[Lemma 4.4]{MOO18}, we have $\mathscr{K}_{i,j,a}= \mathscr{L}_{i,j,a}$, for all $i \in F_{2,a}^{(q)}\cup L_{2,a}^{(q)}$ and $0 \leq j < \ell $. Of course, $\mathscr{K}_{i,j,a}$ is a finite chain ring with invariants $(p, e_2, r \kappa_{i,j,a}^{(q^\ell)}, k_2, t_2)$ with identity $\epsilon_{i,j,a}$. Moreover, $$ 
\mu(\epsilon_{i,j, a})=\left\{
                        \begin{array}{ll}
                          \epsilon_{i, j, a}, & \hbox{if $i\in F_{2,a}^{(q)}$,} \\
                          \epsilon_{-i,j,a}, & \hbox{if $i\in L_{2,a}^{(q)}$.}
                        \end{array}
                      \right.
$$

\begin{lemma}[Lemma 4.2 in \cite{MOO18}]\label{lem:idempotent-R}
 The set of all pairwise orthogonal primitive central idempotent elements of $\mathscr{S}_a$ is $$\{\epsilon_{i,j,a} \mid i \in F_{2,a}^{(q)}\cup L_{2,a}^{(q)} \text{ and }0 \leq   j <\ell\}.$$ Moreover,
 $$\mathscr{S}_a  = \left(\bigoplus\limits_{i\in F_{1,a}^{(q)}}\mathscr{L}_{i,a}\right)\oplus\left(\bigoplus\limits_{i\in L_{1,a}^{(q)}}\mathscr{L}_{i,a}\right)\oplus\left(\bigoplus\limits_{i\in F_{2,a}^{(q)}}\mathscr{L}_{i,a}\right)\oplus\left(\bigoplus\limits_{i\in L_{2,a}^{(q)}}\mathscr{L}_{i,a}\right)$$ and $$\mathscr{L}_{i,a} = \mathscr{L}_{i,0,a} \oplus \mathscr{L}_{i,1,a} \oplus \cdots \oplus \mathscr{L}_{i,\ell-1,a},$$ where  $\epsilon_{i,0,a}+ \epsilon_{i,1,a} + \cdots + \epsilon_{i,\ell-1,a} = \epsilon_{i,a}$ and $\epsilon_{i,a}\epsilon_{h,j,a} = 0$ when $i \neq h$, for all $(h,i)\in Z_a^{(q)}\times(F_{2,a}^{(q)}\cup L_{2,a}^{(q)})$,  and $0 \leq j <\ell$. 
\end{lemma}

 \begin{cor}\label{cor:idempotent-S} 
 Let 
 $$\mathfrak{I}_i^{(q)}=\prod\limits_{a=1}^{\flat}(F_{i,a}^{(q)}\cup L_{i,a}^{(q)}),\text{ where $i\in\{1,2\}$, and }\mathfrak{U}(\flat, \ell)=\{0,1,\ldots,\ell-1\}^{\flat-1}.$$  
  Let 
$
\epsilon_{\mathbf{x},\mathbf{j}} = \epsilon_{{x_1},j_1,1} \cdots \epsilon_{{x_\flat},j_\flat,\flat},
$ for $\mathbf{x}=({x_1},\ldots, {x_\flat})\in \mathfrak{I}_2^{(q)}$ and $\mathbf{j}=(j_1,\ldots, j_\flat)\in\mathfrak{U}(\flat, \ell)$.
Then 
$$
\left\{\epsilon_{\mathbf{x},\mathbf{j}} \mid (\mathbf{x}, \mathbf{j})\in \mathfrak{I}_2^{(q)}\times\mathfrak{U}(\flat, \ell)\right\}
$$
is the set of pairwise orthogonal primitive central idempotent elements of the group ring $\mathscr{S}$, and  \begin{align}\label{deco_S}
                                                                                                  \mathscr{S} =\left(\bigoplus\limits_{\mathbf{x}\in \mathfrak{I}_1^{(q)}}  \mathscr{L}_{ \mathbf{x}}\right)\oplus\left(\bigoplus\limits_{(\mathbf{x}, \mathbf{j})\in \mathfrak{I}_2^{(q)}\times\mathfrak{U}(\flat, \ell)} \mathscr{L}_{\mathbf{x},\mathbf{j}}\right)\,,
                                                                                               \end{align} 
where $\mathscr{L}_{\mathbf{x}}=\mathscr{S}\epsilon_{{\mathbf{x}}}$ and $\mathscr{L}_{\mathbf{x},\mathbf{j}}=\mathscr{S}\epsilon_{\mathbf{x},\mathbf{j}}$.
\end{cor}
Since $\mathscr{L}_{\mathbf{x}} = \bigoplus\limits_{j=0}^{\ell-1} \beta_j \mathscr{K}_{\mathbf{x}}$, from Equation~\eqref{deco_S}, we have the following result that generalizes the result stated for cyclic codes in \cite{MOO18}. 

\begin{theorem}\label{thm3.10} Consider the notations given above, and let  $G$ be an Abelian group such that $\gcd(|G|, p) = 1$.  Any  $S|R$-additive $G$-code $C$ is of the form 
\begin{align}\label{deco_C} C=\sum\limits_{j=0}^{\ell-1}\sum\limits_{\mathbf{x}\in\mathfrak{I}_1^{(q)}}\gamma^{a_{\mathbf{x},j}}\beta_j  \delta_{{\mathbf{x},j}}\mathscr{K}_{\mathbf{x}} +\sum\limits_{(\mathbf{x}, \mathbf{j})\in\mathfrak{I}_2^{(q)}\times\mathfrak{U}(\flat, \ell)} \gamma^{b_{\mathbf{x}, \mathbf{j}}}\omega_{\mathbf{x}, \mathbf{j}}\mathscr{L}_{\mathbf{x}, \mathbf{j}}\,,
\end{align} where $0\leq a_{\mathbf{x},j}, b_{\mathbf{x}, \mathbf{j}}<s$ and $\{\delta_{{\mathbf{x},j}},\omega_{\mathbf{x}, \mathbf{j}}\}\subseteq\{0,1\}$.
\end{theorem} 

\section{Left additive group codes and their duals}\label{sec:dual}

In this section,  $S|R$ will  be an extension of finite chain rings 
such that $S = R[\alpha, \gamma]$, where $S|R_{\alpha}$ is an Eisenstein extension, 
and $S|R_{\gamma}$ is a Galois extension of $\ell$. Let $\underline{\beta} = (\beta_0, \dots, \beta_{\ell-1})$ be an ordered basis of the free $R$-module $R_{\alpha}$, and  $\underline{\beta}^* = (\beta_0^*,  \dots, \beta_{\ell-1}^*)$ the dual basis of $\underline{\beta}$. If $G$ is a finite group,  we will denote by $\mathscr{R}_\alpha,$\label{mathscr{R}_alpha} and $\mathscr{R}_\gamma$\label{mathscr{R}_gamma}  the group  rings ${R}_\alpha [G]$ and ${R}_\gamma [G]$, respectively. As in the previous section, we denote $\mathscr{R}=R[G]$ and $\mathscr{S}=S[G]=R[\alpha, \gamma][G]$.


\subsection{$\texorpdfstring{\mathscr{R}\overline{\mathscr{R}}}{R R}$-ary images of left additive group codes}

The \emph{extended trace map} $\Tr:\mathscr{R}_\alpha\rightarrow\mathscr{R}$ is defined $R$-linearly by
\begin{equation}
    \Tr\left(\sum_{g \in G} a_g g\right) = \sum_{g \in G} \Tr(a_g) g,
\end{equation}
where $\Tr : R_{\alpha} \to R$ is the trace map of the Galois extension $S|R_{\alpha}$. Note that the extended trace $\Tr : \mathscr{R}_\alpha \to \mathscr{R}$ is an $\mathscr{R}$-bimodule epimorphism. Therefore $$\Tr(r' x r) = r' \Tr(x)  r,\quad \forall x \in \mathscr{R}_\alpha,\hbox{ and }(r, r') \in \mathscr{R}^2.$$ On the other hand, $\underline{\beta}$ and $\underline{\beta}^*$ are also ordered bases of the free $R_{\gamma}$-module $S$ and the free $\mathscr{R}_\gamma$-module $\mathscr{S}$. The maps $\mathrm{d}_{i,\underline{\beta}}$ and $\mathrm{d}_{\underline{\beta}}$ defined in  Equations~\eqref{eq:d_i_alpha} and \eqref{eq:d_alpha}, respectively, can be extended to the group ring $\mathscr{S}$ as follows:

\begin{align}
\mathrm{D}_{i,\underline{\beta}} : \mathscr{S} &\longrightarrow \mathscr{R}_\gamma, \quad \sum_{g \in G} a_g g \longmapsto \sum_{g \in G} \mathrm{d}_{i,\underline{\beta}}(a_g) g, \label{eq:D_i_alpha} \\
\mathrm{D}_{\underline{\beta}} : \mathscr{S} &\longrightarrow (\mathscr{R}_\gamma)^\ell, \quad x \longmapsto (\mathrm{D}_{0,\underline{\beta}}(x), \dots, \mathrm{D}_{\ell-1,\underline{\beta}}(x)). \label{eq:D_alpha}
\end{align}

  Note that in Equations~\eqref{eq:D_i_alpha} and \eqref{eq:D_alpha}  we have that   $\mathrm{D}_{i,\underline{\beta}}$ is an $\mathscr{R}$-module epimorphism, 
and $\mathrm{D}_{\underline{\beta}}$ is an $\mathscr{R}$-module isomorphism.

The map $\psi$ defined in Equation~\eqref{psi} can be extended to the direct product of group rings as follows: 
{\small \begin{align}\label{Psi1}
\begin{array}{cccc}
  \Psi: & \prod\limits_{i=0}^{k-1} \gamma^i\mathscr{R}_\alpha & \longrightarrow & (\mathscr{R}_\alpha)^t \times (\overline{\mathscr{R}}_{\overline{\alpha}})^{k-t} \\
    & (x_0,\cdots, x_{k-1}) & \longmapsto & \left( \Psi_0(x_0), \dots, \Psi_{t-1}(x_{t-1}) \mid \overline{\Psi}_t(x_t), \dots, \overline{\Psi}_{k-1}(x_{k-1}) \right), 
\end{array}
\end{align}}%
where 
\begin{align}
\Psi_i \colon \gamma^i \mathscr{R}_\alpha &\longrightarrow \mathscr{R}_\alpha, & \gamma^i x &\longmapsto x,\label{Psi_i} \\
\overline{\Psi}_j \colon \gamma^j \mathscr{R}_\alpha &\longrightarrow \overline{\mathscr{R}}_{\overline{\alpha}}, & \gamma^j x &\longmapsto x + \gamma_1^{s_1-1}\mathscr{R},\label{Psi_j}
\end{align}
for $0 \leq i < t \leq j < k$ and $\overline{\mathscr{R}}_{\overline{\alpha}} \coloneqq \mathscr{R}_\alpha/\gamma_1^{s_1-1}\mathscr{R}_\alpha$. Note that, since $\{1, \gamma, \dots, \gamma^{k-1}\}$ is an $R_{\alpha}$-independent generating set of $S$, the set $\{1, \gamma, \dots, \gamma^{k-1}\}$ is also an $\mathscr{R}_\alpha$-independent generating set of $\mathscr{S}$.
Therefore there are $\mathscr{R}_\alpha$-module epimorphisms
\begin{align}\label{phi_i}
  \begin{array}{cccc}
\phi_i : & \mathscr{S} & \longrightarrow & \gamma^i \mathscr{R}_\alpha \\
  & x  & \longmapsto  &\phi_i(x),
  \end{array}
\end{align} 
for $0\leq i< k$ such that 
\begin{align}\label{phi}
\begin{array}{cccc}
\phi : & \mathscr{S} & \longrightarrow & \prod\limits_{i=0}^{k-1}\gamma^i \mathscr{R}_\alpha \\
  & x  & \longmapsto  &(\phi_0(x),\ldots, \phi_{k-1}(x))
  \end{array}
\end{align}
is an $\mathscr{R}_\alpha$-module isomorphism. 
Therefore, \begin{align}\label{upsilon}
              \Lambda=\Psi\circ\phi : \mathscr{S}\longrightarrow (\mathscr{R}_\alpha)^t\times(\overline{\mathscr{R}}_{\overline{\alpha}})^{k-t}
           \end{align}
is an $\mathscr{R}_\alpha$-module isomorphism, and the other hand,  
for all $0 \leq i < t \leq j < k$, the maps \begin{align}\label{upsilons}\Lambda_i=\Psi_i\circ\phi_i: \mathscr{S}\longrightarrow \mathscr{R}_\alpha\text{ and }\overline{\Lambda}_j=\overline{\Psi}_j\circ\phi_j : \mathscr{S}\longrightarrow \overline{\mathscr{R}}_{\overline{\alpha}}\end{align}
are $\mathscr{R}_\alpha$-module epimorphisms. The commutative diagram in Equation~\eqref{diag} leads to the following one
\begin{align}
\label{diag1}
\begin{array}{ccc}
\mathscr{S} & \xrightarrow{\quad\mathrm{D}_{\underline{\beta}}\quad} & (\mathscr{R}_\gamma)^\ell \\
\Big\downarrow{\scriptstyle\Lambda} &  & \Big\downarrow{\scriptstyle \tau \circ \Lambda^{\times \ell}} \\[2ex]
(\mathscr{R}_\alpha)^t \times (\overline{\mathscr{R}}_{\overline{\alpha}})^{k-t} & \xrightarrow{\quad \mathrm{D}_{\underline{\beta}}^{\times t} \times \overline{\mathrm{D}}_{\underline{\beta}}^{\times (k-t)} \quad} & \mathscr{R}^{t\ell} \times \overline{\mathscr{R}}^{(k-t)\ell}.
\end{array}
\end{align}
\begin{remark}\label{reGal}
    In the case the extension $S|R$ is Galois, then $k=t=1$, and
    $ \Lambda$ is the identity map. On the other hand, if the extension $S|R$ is Eisenstein, then $\ell=1$ and $\mathrm D_{\underline{\beta}}$ is the identity map. 
\end{remark}

\begin{remark} If $k = t = 1$, then the extension $S|R$ is Galois, and hence the map $\Lambda$ in Equation~\eqref{diag1} is the identity. On the other hand, if $\ell = 1$, then the extension $S|R$ is Eisenstein, and thus the map $\mathrm{D}_{\underline{\beta}}$ in Equation~\eqref{diag1} is the identity. \end{remark}

Commutative diagram \ref{diag1} leads us to obtain the $\mathscr{R}$-module isomorphism
\begin{align}
\begin{array}{cccc}
  \Theta \colon & \mathscr{S}  & \longrightarrow &  \mathscr{R}^{t\ell} \times \overline{\mathscr{R}}^{(k-t)\ell} \\
    & x & \longmapsto &  \big(\Theta_{ 0}(x), \ldots, \Theta_{\ell t-1}(x) \mid \overline{\Theta}_{\ell t}(x), \ldots, \overline{\Theta}_{\ell k-1}(x)\big), \label{Lamb} 
\end{array}
\end{align}
defined by
$
\Theta = \left(\mathrm{D}_{\underline{\beta}}^{\times t} \times \overline{\mathrm{D}}_{\underline{\beta}}^{\times (k-t)}\right) \circ \Lambda,$
where
\begin{align}\label{lambs}\Theta_{i\ell + u}= \mathrm{D}_{u,\underline{\beta}}\circ \Lambda_i : \mathscr{S}\longrightarrow\mathscr{R},
\text{
 and
}
\overline{\Theta}_{j\ell+u} = \overline{\mathrm{D}}_{v,\underline{\beta}}\circ\overline{\Lambda}_{j} 
: \mathscr{S}\longrightarrow \overline{\mathscr{R}},
\end{align} for all indices satisfying $0 \le i < t \le j < k$, and $0 \le u , v< \ell$. 
  By Remark~\ref{rem:Rset}, we have the decomposition
\begin{align}
    \mathscr{S}=  \left(\bigoplus_{u=0}^{\ell-1} 
    \bigoplus_{i=0}^{t-1} \mathscr{R}\alpha_u\gamma^i \right)
    \oplus \left(\bigoplus_{v=0}^{\ell-1}
    \bigoplus_{j=t}^{k-1}  \mathscr{R}\alpha_v\gamma^j 
\right) 
\end{align}
that induces 
\begin{align}\label{exp_Theta}
\begin{array}{cccc}
\Theta : & \mathscr{S} & \longrightarrow & \mathscr{R}^{t\ell}\times\overline{\mathscr{R}}^{(k-t)\ell} \\
  & \left(\sum\limits_{u=0}^{\ell-1} 
    \sum\limits_{i=0}^{t-1} x_{i,u}\alpha_u\gamma^i \right)
    +
   \left(\sum\limits_{v=0}^{\ell-1}  \sum\limits_{j=t}^{k-1}  x_{j,u}\alpha_v\gamma^j 
\right)   & \longmapsto  & \left((x_{i,u})_{i,u}\;|\; (\overline{x_{j,v}})_{j,v}\right).
  \end{array} \end{align} 
With the module structure defined in~\eqref{Rmod}, we recall that the left $\mathscr{R}$-submodules of the module 
$\mathscr{R}^{\ell t} \times \overline{\mathscr{R}}^{\ell(k-t)}$ are called \emph{$\mathscr{R}\overline{\mathscr{R}}$-linear codes} 
of block-length $(\ell t, \ell(k-t))$, or simply \emph{mixed-alphabet codes} (see, e.g.,~\cite{BDFM25}). 

The $\mathscr{R}$-module isomorphism $\Theta \colon \mathscr{S} \rightarrow \mathscr{R}^{\ell t} \times \overline{\mathscr{R}}^{\ell(k-t)}$ 
yields the following characterization.

\begin{prop}\label{p1}
Let $\emptyset \neq C \subseteq \mathscr{S}$. Then $C$ is  {a left $S|R$-additive $\mathrm{G}$-code} if and only if $\Theta(C)$ is a left $\mathscr{R}\overline{\mathscr{R}}$-linear code 
of block-length $(\ell t, \ell(k-t))$. 
\end{prop}

\begin{proof}
    The proof of this proposition follows immediately from the properties of $\Theta$ and the definition of additive group codes.
\end{proof}

We will call the $\mathscr{R}\overline{\mathscr{R}}$-linear code $\Theta(C)$
of block-length $(\ell t, \ell(k-t))$ the $\mathscr{R}\overline{\mathscr{R}}$-ary image of {the left $S|R$-additive $\mathrm{G}$-code $C$}.

\begin{remark} \label{rem3:ideal_decomp}
Let $C$ be {a left $S|R$-additive $\mathrm{G}$-code}. Then $\Theta_i(C)$ is an ideal of $\mathscr{R}$ and $\overline{\Theta}_j(C)$ is an ideal of $\overline{\mathscr{R}}$ for $0 \leq i < \ell t \leq j < \ell k$. Moreover,
\begin{align}
\Theta(C) = \left( \prod_{i=0}^{\ell t-1} \Theta_i(C) \right) \times \left( \prod_{j=\ell t}^{\ell k-1} \overline{\Theta}_j(C) \right). \label{Eq:lam_prod}
\end{align}
\end{remark}

\begin{example}\label{Ex-ary}
Consider $S = \mathbb{Z}_9[\alpha, \gamma]$, where $\alpha$ and $\gamma$ are as defined in Example~\ref{ex_S.R}, and the group $\mathbb{S}_3$ with the presentation given in Example~\ref{ExG-code1}.  Equation \eqref{exp_Theta} allows us to obtain
$$
\begin{array}{cccc}
\Theta : & \mathscr{S} & \longrightarrow & \mathscr{R}^2\times\overline{\mathscr{R}}^2 \\
  & x_{0,0}+x_{0,1}\alpha+x_{1,0}\gamma+x_{1,1}\alpha\gamma  & \longmapsto  & \left(x_{0,0},\; x_{0,1}\;|\; \overline{x_{1,0}},\; \overline{x_{1,1}}\right),
  \end{array} $$ 
where $\mathscr{R}:=\mathbb{Z}_{9}[\mathbb{S}_3],\overline{\mathscr{R}}:=\mathbb{Z}_{3}[\mathbb{S}_3]$, and $\overline{x}$ denotes reduction modulo $3$.  
For $\beta = 4 + \alpha$, the code
  $$C= \mathscr{R}\beta(1-\varrho_2)+\mathscr{R}\gamma \beta(1-\varrho_1)$$ is a left $S|\mathbb{Z}_9$-additive $\mathbb{S}_3$-code. It is easy to check that $$\mathscr{R}\beta (1-\varrho_2)\cap\mathscr{R}\gamma \beta(1-\varrho_1)=\{\mathbf{0}\}.$$ Using the dual basis $(2\alpha, 2 + \alpha)$ of $(1, \alpha)$, we have  $\Tr(\beta(2\alpha))=4$ and $\Tr(\beta(2+\alpha))=1$.
Henceforth, $\Theta(C)=(\mathscr{R}(1-\varrho_2))^2\times (\overline{\mathscr{R}}(1-\varrho_1))^2.$
\end{example}

\subsection{Duals of left additive group codes}\label{ssec:dual}

In \cite{DOUGHERTY2021}, the authors defined several dualities for $\mathbb{F}_{q^m}|\mathbb{F}_q$-additive $G$-codes. They prove that, for such a code $\mathcal{C}$, its dual (under the  duality they consider) $\mathcal{C}^\perp$ is also a  an $\mathbb{F}_{q^m}|\mathbb{F}_q$-additive $G$-code.  

Consider now, for example, the ring extension $S = R[\alpha, \gamma]$ where $R = \mathbb{Z}_9$, and $\gamma^2 - 3 = 3\gamma = 0$ and $\alpha^2 = 5\alpha + 1$. For the $S|R$-additive code $C = \gamma R + \alpha R$, the Euclidean dual is given by:
\[
C^{\perp_E} 
= (\gamma R)^{\perp_E} \cap (\alpha R)^{\perp_E} 
= \mathfrak{m}_S^2 \cap \{0\} 
= \{0\}.
\]
 Note that $w\in (\alpha R)^{\perp_E}$ if $\alpha rw=0$ for all $r\in R$, i.e., $\alpha w=0$. Since $\alpha$ is a unit in $S$, then $w=0$.
Thus, 
 $
(C^{\perp_E})^{\perp_E} = \{0\}^{\perp_E} = S \neq C
$, and the double dual property is not fulfilled.

Therefore, the Euclidean dual fails to preserve the code structure and violates the double-dual property. The goal of this subsection is to define a \emph{natural duality} on the $R$-module $\mathscr{S}$ such  that
\begin{enumerate}
    \item it preserves the structure of $S|R$-additive $G$-codes,
    \item it ensures that the dual of any such code is also an $S|R$-additive $G$-code,
    \item it satisfies the double-dual property: $(C^\perp)^\perp = C$,
    \item it yields the standard Euclidean duality when $S = R$.
\end{enumerate}
This duality will be induced by a symmetric non-degenerate bilinear form on the group ring $\mathscr{S}$.  

The $R_\alpha$-module $(R_\alpha)^{t} \times (\overline{R}_{\overline{\alpha}})^{k-t}$ arises from the Eisenstein extension $S|R_\alpha$ of block-degree $(k,t)$ under  the image of a module by the map $\xi$ defined in Equation \eqref{xi}, and the $R_\alpha$-submodules of the $R_\alpha$-module $(R_\alpha)^{t} \times (\overline{R}_{\overline{\alpha}})^{k-t}$ are called $R_\alpha\overline{R}_{\overline{\alpha}}$-linear codes of block-length $(t,k-t)$. In \cite{BDFM25, JS24c}, the Euclidean duality of $R_\alpha\overline{R}_{\overline{\alpha}}$-linear codes has been studied. Given two elements $ \mathbf{v}, \mathbf{w}\in  (R_\alpha)^{t} \times (\overline{R}_{\overline{\alpha}})^{k-t}$ ,   their \emph{Euclidean inner product} is defined as
\begin{align}\label{[]_E}
\left[\,\mathbf{v},\mathbf{w}\right]_E=\sum\limits_{i=0}^{ t-1}v_iw_i+\chi\left(\sum\limits_{j= t}^{ k-1}\overline{v}_j\overline{w}_j\right),
\end{align}
where
\begin{align}\label{chi}
\begin{array}{cccc}
  \chi: & \overline{R}_{\overline{\alpha}} & \longrightarrow & R_{\alpha}\mathfrak{m}_R \\
    & \displaystyle\sum_{i=0}^{s-2} x_{i} \, \overline{\gamma_1}^i & \longmapsto & \gamma_1  \displaystyle\sum_{i=0}^{s-2} \iota(x_{i})\, \gamma_1^i,
\end{array}
\end{align}
and the map $\iota : \Gamma(\overline{R}) \to \Gamma(R)$ \label{iota} satisfies $\iota(0) = 0$ with its restriction to $\Gamma(\overline{R})^*$ being a group isomorphism. Note that by \cite[Lemma 1]{BDFM25}, $\chi$ is an $R_{\alpha}$-module isomorphism.

Using the $\mathscr{R}_{\alpha}$-isomorphism $\Lambda : \mathscr{S} \longrightarrow (\mathscr{R}_{\alpha})^t \times (\overline{\mathscr{R}}_{\overline{\alpha}})^{k-t}$ defined in Equation \eqref{upsilon}, we endow the $\mathscr{R}_{\alpha}$-module $\mathscr{S}$ with the \emph{$\Lambda$-Euclidean inner product} defined by
\begin{align}\label{inner_Lambda}
\mathbf{v}\ast_\Lambda \mathbf{w}=[\Lambda(\mathbf{v}),\Lambda(\mathbf{w})]_E, \hbox{ where } (\mathbf{v},\mathbf{w}) \in \mathscr{S}^2.
\end{align}
Note that if $k=t=1$, we have $\mathbf{v} \ast_\Lambda \mathbf{w}=\mathbf{v}\cdot\mathbf{w}${, where $\mathbf{v}\cdot\mathbf{w}$ is the standard Euclidean inner product over $\mathscr{S}$ induced by the identification provided in Equation~\eqref{PhiO}.}
Equations \eqref{inner} and \eqref{eq:dot_product} show that the $R_{\alpha}$-bilinear extension to $\mathscr{R}_{\alpha}$ corresponds to the usual Euclidean inner product on $(R_{\alpha})^n$. Therefore, for two elements  $\mathbf{v},\mathbf{w} \in \mathscr{S}$, their inner product is given by       
 \begin{align}\label{eq:upsi_inner}
\mathbf{v} \ast \mathbf{w} = \coeffId{\mathbf{v}\ast_\Lambda \mathbf{w}}= \sum_{i=0}^{t-1}\coeffId{\Lambda_i(\mathbf{v})\Lambda_i(\mathbf{w})} + \chi\left(\sum_{j=t}^{k-1} \coeffId{ \overline{\Lambda}_j(\mathbf{v})\overline{\Lambda}_j(\mathbf{w})}\right){\in R_{\alpha} },
\end{align} 
where $\coeffId{\cdot} \colon \mathscr{R}_{\alpha} \to R_{\alpha}$  is the $R_{\alpha}$-module epimorphism defined in Equation~\eqref{coef}.   Moreover, we have 
    \begin{align*}
        \coeffId{\Lambda_i(\mathbf{v}) \Lambda_i(\mathbf{w})} 
    &= \left\langle \Lambda_i(\mathbf{v}), \mu(\Lambda_i(\mathbf{w})) \right\rangle\in R_\alpha,\\
    \coeffId{\overline{\Lambda}_j(\mathbf{v}) \overline{\Lambda}_j(\mathbf{w})} 
    &= \left\langle \overline{\Lambda}_j(\mathbf{v}), \mu(\overline{\Lambda}_j(\mathbf{w})) \right\rangle\in \overline{R}_{\overline{\alpha},}
    \end{align*}
    for all $0\leq i<t\leq j  < k$.    
Finally, the trace map $\Tr: R_\alpha \rightarrow R$, defined for the Galois extension $R_\alpha|R$  allows us to define the \emph{$\Theta$–Euclidean inner product} $\circledast$ on $\mathscr{S}$ as
\begin{align}\label{eq:inner}
\mathbf{v} \circledast \mathbf{w} = \Tr\left( \mathbf{v} \ast \mathbf{w} \right). 
\end{align}
It is easy to check that   $\circledast$  is a symmetric, non-degenerate $R$-bilinear form on the group ring $\mathscr{S}$.
\begin{definition}
    Let $C$ be a left $S|R$-additive group code. The \emph{$\circledast$-dual} of $C$  is the set
\begin{align}\label{dual}
C^{\perp_{\circledast}} = \left\{ \mathbf{w} \in \mathscr{S} \mid\mathbf{c} \circledast \mathbf{w} = 0 \text{ for all } \mathbf{c} \in C \right\}.
\end{align}
\end{definition}

{\begin{lemma}\label{m28} Let $(\mathbf{v}, \mathbf{w}) \in \mathscr{S}^2$ and $g \in G$. Suppose that $C$ is a left $S|R$-additive $G$-code. Then:
\begin{enumerate}
    \item $ g \mathbf{v} \circledast \mathbf{w} = \mathbf{v} \circledast g^{-1} \mathbf{w} $.
    \item The code $C^{\perp_\circledast}$ is also a left $S|R$-additive $G$-code.
\end{enumerate}
\end{lemma}
\begin{proof}
Assume that $\Lambda_i(\mathbf{v}) = \sum_{h \in G} a_h h$. Then
\begin{align*}
\Lambda_i(g \mathbf{v}) \Lambda_i(\mathbf{w}) &= \left( \sum_{h \in G} a_h (g h) \right) \left( \sum_{k \in G} b_k k \right),\\
\Lambda_i(\mathbf{v}) \Lambda_i(g^{-1} \mathbf{w}) &= \left( \sum_{h \in G} a_h h \right) \left( \sum_{k \in G} b_k (g^{-1} k) \right).
\end{align*}
Therefore, 
\begin{align*}
 \coeffId{\Lambda_i(g \mathbf{v}) \Lambda_i(\mathbf{w})} &= \sum_{\substack{h, k \in G \\ g h k = g_1}} a_h b_k = \sum_{h \in G} a_h b_{h^{-1} g^{-1}};\\
 \coeffId{\Lambda_i(\mathbf{v}) \Lambda_i(g^{-1} \mathbf{w})} &= \sum_{\substack{h, k \in G \\ h (g^{-1} k) = g_1}} a_h b_k = \sum_{h \in G} a_h b_{g h^{-1}}.
\end{align*}
Reindexing $h' = h^{-1}$, one has $$\sum_{h \in G} a_h b_{g h^{-1}} = \sum_{h' \in G} a_{(h')^{-1}} b_{g h'} = \sum_{h \in G} a_h b_{h^{-1} g^{-1}},$$
and therefore
$$
\coeffId{\Lambda_i(g \mathbf{v}) \Lambda_i(\mathbf{w})} = \coeffId{\Lambda_i(\mathbf{v}) \Lambda_i(g^{-1} \mathbf{w})}.
$$
Similarly,  for $\overline{\Lambda}_j$, since $\chi$ is an isomorphism, we obtain
$$
\coeffId{\overline{\Lambda}_j(g \mathbf{v}) \overline{\Lambda}_j(\mathbf{w})} = \coeffId{\overline{\Lambda}_j(\mathbf{v}) \overline{\Lambda}_j(g^{-1} \mathbf{w})}.
$$
Now, the following equations prove the first statement:
\begin{align*}
g \mathbf{v} \circledast \mathbf{w} &= \Tr\left( \sum_{i=0}^{t-1} \coeffId{\Lambda_i(g \mathbf{v}) \Lambda_i(\mathbf{w})} + \chi\left( \sum_{j=t}^{k-1} \coeffId{\overline{\Lambda}_j(g \mathbf{v}) \overline{\Lambda}_j(\mathbf{w})} \right) \right), \\
\mathbf{v} \circledast g^{-1} \mathbf{w}  &= \Tr\left( \sum_{i=0}^{t-1} \coeffId{\Lambda_i(\mathbf{v}) \Lambda_i(g^{-1} \mathbf{w})} + \chi\left( \sum_{j=t}^{k-1} \coeffId{\overline{\Lambda}_j(\mathbf{v}) \overline{\Lambda}_j(g^{-1} \mathbf{w})} \right) \right).
\end{align*}
The second statement is an immediate consequence of the first one.
\end{proof}}

The following result establishes a relationship between the Euclidean inner product $[\cdot,\cdot]_E$ on $\mathscr{R}^{t\ell} \times \overline{\mathscr{R}}^{(k-t)\ell}$ and the inner product $\circledast$ on the group ring $\mathscr{S}$.

\begin{lemma}\label{lem:inner_rel2}
Let $(\mathbf{v}, \mathbf{w}) \in \mathscr{S}^2$. Then
\begin{align}
[\Theta(\mathbf{v}), \Theta(\mathbf{w})]_E =  \sum\limits_{g\in G}(g^{-1}\mathbf{v} \circledast\mathbf{w}). \label{eq:rel_inner3}
\end{align}
\end{lemma} 

\begin{proof} Let $(\mathbf{v}, \mathbf{w}) \in \mathscr{S}^2$. Then $$\Lambda(\mathbf{v}) =(v_0,\ldots, v_{t-1}\,|\,\overline{v}_t,\ldots, \overline{v}_{k-1})\text{
and }\Lambda(\mathbf{w})= (w_0,\ldots, w_{t-1}\,|\,\overline{w}_t,\ldots, \overline{w}_{k-1})$$ are elements of $(\mathscr{R}_{\alpha})^{t\ell} \times (\overline{\mathscr{R}}_{\overline{\alpha}})^{(k-t)\ell}$. 
Therefore, we obtain
\begin{align*}
\Theta(\mathbf{v}) &= 
\bigl(\mathrm{D}_{\underline{\beta}}^{\times t} 
\times \overline{\mathrm{D}}_{\underline{\beta}}^{\times (k-t)}\bigr)
\circ \Lambda(\mathbf{v}) \\
&= \bigl(
\mathrm{D}_{\underline{\beta}}(v_0), \ldots, \mathrm{D}_{\underline{\beta}}(v_{t-1})
\,\mid\,
\overline{\mathrm{D}}_{\underline{\beta}}(\overline{v}_t), \ldots,
\overline{\mathrm{D}}_{\underline{\beta}}(\overline{v}_{k-1})
\bigr), \\[6pt]
\Theta(\mathbf{w}) &= 
\bigl(\mathrm{D}_{\underline{\beta}}^{\times t} 
\times \overline{\mathrm{D}}_{\underline{\beta}}^{\times (k-t)}\bigr)
\circ \Lambda(\mathbf{w}) \\
&= \bigl(
\mathrm{D}_{\underline{\beta}}(w_0), \ldots, \mathrm{D}_{\underline{\beta}}(w_{t-1})
\,\mid\,
\overline{\mathrm{D}}_{\underline{\beta}}(\overline{w}_t), \ldots,
\overline{\mathrm{D}}_{\underline{\beta}}(\overline{w}_{k-1})
\bigr),
\end{align*}
where 
\begin{align*}
\mathrm{D}_{\underline{\beta}}(v_i) &=(v_{0,i},\ldots, v_{\ell-1,i}),\;\overline{\mathrm{D}}_{\underline{\beta}}(\overline{v}_j)=(\overline{v}_{0,j},\ldots, \overline{v}_{\ell-1,j}),\\
\mathrm{D}_{\underline{\beta}}(w_i) &=(w_{0,i},\ldots, w_{\ell-1,i}),\;\overline{\mathrm{D}}_{\underline{\beta}}(\overline{w}_j)=(\overline{w}_{0,j},\ldots, \overline{w}_{\ell-1,j}),\end{align*} for all $0\leq i< t\leq j< k$. 
On the one hand, applying \eqref{coef}, we have 
\begin{eqnarray*} 
[\Theta(\mathbf{v}), \Theta(\mathbf{w})]_E &=&\sum_{x=0}^{\ell-1}\left(\sum_{i=0}^{t-1}v_{x,i}w_{x,i} + \chi\left(\sum_{j= t}^{k-1} \overline{v}_{x,j}\overline{w}_{x,j}\right)\right)\\
     &=&\sum_{g\in G}\left(\sum_{x=0}^{\ell-1}\left(\sum_{i=0}^{t-1}\coeffId{g^{-1}v_{x,i}w_{x,i}} + \chi\left(\sum_{j= t}^{k-1} \coeffId{g^{-1}\overline{v}_{x,j}\overline{w}_{x,j}}\right)\right)\right)g.
\end{eqnarray*}
Moreover,
\begin{eqnarray*}\Lambda(\mathbf{v}) =\big(\Lambda_0(\mathbf{v}), \ldots, \Lambda_{t-1}(\mathbf{v}) \mid \overline{\Lambda}_{t}(\mathbf{v}), \ldots, \overline{\Lambda}_{ k-1}(\mathbf{v})\big),\\
\Lambda(\mathbf{w}) =\big(\Lambda_0(\mathbf{w}), \ldots, \Lambda_{t-1}(\mathbf{w}) \mid \overline{\Lambda}_{t}(x), \ldots, \overline{\Lambda}_{ k-1}(\mathbf{w})\big), 
\end{eqnarray*}
where 
\begin{eqnarray*} \Lambda_i(\mathbf{v})&=&\sum\limits_{x=0}^{\ell-1}v_{x,i}\beta_x,\;\overline{\Lambda}_j(\mathbf{v})=\sum\limits_{x=0}^{\ell-1}\overline{v}_{x,j}\overline{\beta}_x,\\
\Lambda_i(\mathbf{w})&=&\sum\limits_{x=0}^{\ell-1}w_{x,i}\beta_x,\;\overline{\Lambda}_j(\mathbf{w})=\sum\limits_{x=0}^{\ell-1}\overline{w}_{x,j}\overline{\beta}_x,
\end{eqnarray*} for all $0\leq i< t\leq j< k$. Therefore
\begin{eqnarray*} 
g^{-1}\mathbf{v} \circledast \mathbf{w}  &=& \Tr(g^{-1}\mathbf{v}\ast \mathbf{w})\\
    &=&  \sum_{i=0}^{t-1}\Tr\left(\coeffId{g^{-1}\Lambda_i(\mathbf{v})\Lambda_i(\mathbf{w})}\right) +\sum_{j=t}^{k-1}\chi\left(\Tr\left(\coeffId{g^{-1} \overline{\Lambda}_j(\mathbf{v})\overline{\Lambda}_j(\mathbf{w})}\right)\right)\\
    &=& \sum_{i=0}^{t-1}\Tr\left(\coeffId{\left(\sum\limits_{x=0}^{\ell-1}g^{-1}v_{x,i}\beta_x\right)\left( \sum\limits_{y=0}^{\ell-1}w_{y,i}\beta_y \right)}\right)\\
    & & + \sum_{j=t}^{k-1}\chi\left(\Tr\left(\coeffId{\left(\sum\limits_{x=0}^{\ell-1}g^{-1}\overline{v}_{x,j}\beta_x\right)\left( \sum\limits_{y=0}^{\ell-1}\overline{w}_{y,j}\beta_y \right)}\right)\right)\\
    &=& \sum_{x=0}^{\ell-1}\left(\sum_{i=0}^{t-1}\coeffId{g^{-1}v_{x,i}w_{x,i}} + \chi\left(\sum_{j= t}^{k-1} \coeffId{g^{-1}\overline{v}_{x,j}\overline{w}_{x,j}}\right)\right).
\end{eqnarray*}
Hence, $[\Theta(\mathbf{v}), \Theta(\mathbf{w})]_E = \sum\limits_{g\in G}(g^{-1}\mathbf{v} \circledast\mathbf{w})$.
\end{proof}

\begin{theorem}\label{thm4:main}
Let $C$ be a left $S|R$-additive $G$-code. Then
\begin{align}
\Theta(C^{\perp_{\circledast}}) = (\Theta(C))^{\perp_E}. \label{eq:rel_dual3}
\end{align}
\end{theorem}

\begin{proof} On  one hand, we have
$$
   \Theta(C^{\perp_{\circledast}}) = 
    \bigl\{\Theta(\mathbf{w}) \;:\; \mathbf{w} \in C^{\perp_{\circledast}} \bigr\} =
    \bigl\{\Theta(\mathbf{w}) \;:\; (\forall \mathbf{c} \in C)(\mathbf{c} \circledast \mathbf{w} = 0) \bigr\}.
$$
And on the other hand,
$$
    \bigl( \Theta(C) \bigr)^{\perp_E} = 
    \bigl\{\Theta(\mathbf{w}) \;:\; (\forall \mathbf{c} \in C)([\Theta(\mathbf{w}),\Theta(\mathbf{c})]_E = 0) \bigr\}.
$$
Since $C$ is a left $S|R$-additive $G$-code, Lemma~\ref{lem:inner_rel2} implies
$$
    (\forall \mathbf{c} \in C)(\mathbf{c} \circledast \mathbf{w} = 0) 
    \Leftrightarrow (\forall g \in G)(\forall \mathbf{c} \in C)(g^{-1}\mathbf{c} \circledast   \mathbf{w}  = 0)
    \Leftrightarrow [\Theta(\mathbf{c}),\Theta(\mathbf{w})]_E = 0.
$$
Therefore, we conclude that
$
   \Theta(C^{\perp_{\circledast}}) = \bigl( \Theta(C) \bigr)^{\perp_E}.
$
\end{proof}

\begin{cor} Let $C$ be a left $S|R$-additive $G$-code. Then 
    \begin{align}\label{Eq_Iden}(C^{\perp_{\circledast}})^{\perp_{\circledast}} = C\text{  and }|C| \cdot |C^{\perp_{\circledast}}| = |\mathscr{S}|.\end{align}
\end{cor}
 
\begin{proof} If we consider Equation~\eqref{eq:rel_dual3}, we have
\[
\Theta\left((C^{\perp_{\circledast}})^{\perp_{\circledast}}\right) 
= \Theta(C^{\perp_{\circledast}})^{\perp_E} 
= (\Theta(C)^{\perp_E})^{\perp_E} 
= \Theta(C).
\]
Since $\Theta$ is a bijective map, it follows that 
$
(C^{\perp_{\circledast}})^{\perp_{\circledast}} = C.
$
Finally, $|C| \cdot |C^{\perp_{\circledast}}| = |\mathscr{S}|$, since $\mathscr{S}$ is a Frobenius ring and therefore, the sizes condition holds for $\perp_E$.
\end{proof}

\begin{example}
Let $S = \mathbb{Z}_9[\alpha, \gamma]$ be the extension of $\mathbb{Z}_9$  described in Example \ref{ex_S.R}. Then,  $\mathbb{Z}_9[\alpha]|\mathbb{Z}_9$  is a Galois extension of degree $ 2$, and  $S|\mathbb{Z}_9[\alpha]$  is an Eisenstein extension of block-degree $(2, 1)$. Let $\mathbb{S}_3$ be the symmetric group with the presentation in Example~\ref{ExG-code1}. Consider the group rings $\mathscr{R} = \mathbb{Z}_9[\mathbb{S}_3]$, and the code left $S|R$-additive $\mathbb{S}_3$-code  $C$  given in Example \ref{Ex-ary}.  Note that $(1-\varrho_2)(1+\varrho_2)=0=(1-\varrho_1)(1+\varrho_1+\varrho_1^2)$. Therefore $\mathscr{R}(1 - \varrho_2)^{\perp_E} = \mathscr{R}(1 + \varrho_2)$, and $\mathscr{R}(1 - \varrho_1)^{\perp_E} = \mathscr{R}(1 + \varrho_1 + \varrho_1^2)$. Thus $$\Theta(C)^{\perp_E} = (\mathscr{R}(1 + \varrho_2))^2 \times (\overline{\mathscr{R}}(1 + \varrho_1 + \varrho_1^2))^2,$$
and since $\Tr(\beta\beta^*)=0$, we have 
$$
C^{\perp_{\circledast}} = \mathscr{R} \beta^* (1 + \varrho_2) + \mathscr{R} \gamma\beta^* (1 + \varrho_1 + \varrho_1^2),
$$ where $\beta^*=8\alpha+5$. 
\end{example}

In the case of Abelian additive codes, where the Abelian group fulfils $p\nmid |G|$, the $\circledast$-dual code is characterized by the following result.

\begin{cor}\label{cor-dual-abelian} Let $G$ be an Abelian group such that $p\nmid |G|$ and $C$ be a $S|R$-additive $G$-code expressed as in Equation~\eqref{deco_C}. Then   
\begin{align}\label{dual_C} C^{\perp_{\circledast}}=\sum\limits_{j=0}^{\ell-1}\sum\limits_{\mathbf{x}\in\mathfrak{I}(\flat, \mathbf{u})}\gamma^{s-a_{\mathbf{x},j}}\beta^*_j  \delta_{a_{\mathbf{x},j}}\mathscr{K}_{\mathbf{x}} +\sum\limits_{(\mathbf{x}, \mathbf{j})\in\mathfrak{I}_{\mathbf{u}}(\flat, \mathbf{v})\times\mathfrak{U}(\flat, \ell)} \gamma^{s-b_{\mathbf{x}, \mathbf{j}}}\omega_{\mathbf{x}, \mathbf{j}}\mathscr{L}_{\mathbf{x}, \mathbf{j}},
\end{align} where $0\leq a_{\mathbf{x},j}, b_{\mathbf{x}, \mathbf{j}}<s$ and $\{\delta_{a_{\mathbf{x},j}};\omega_{\mathbf{x}, \mathbf{j}}\}\subseteq\{0;1\}$.
\end{cor}

\begin{proof}
 Let the code $C$ be expressed as in Equation~\eqref{deco_C} in Theorem \ref{thm3.10}. Let us denote by $D$ the right-hand side of Equation~\eqref{dual_C}. We will show first that every element in $D$ is orthogonal under the trace inner product to every element in $C$.   
Note that each summand in Equation~\eqref{deco_C} is  of the form
$$
\gamma^{a_{\mathbf{x},j}}\beta_j  \delta_{a_{\mathbf{x},j}} \epsilon_{\mathbf{x}}\mathbf{k}_{\mathbf{x}}+\gamma^{b_{\mathbf{x}, \mathbf{j}}}\omega_{\mathbf{x}, \mathbf{j}}\epsilon_{\mathbf{x},\mathbf{j}}\mathbf{l}_{\mathbf{x},\mathbf{j}},
\hbox{ for some } (\mathbf{k},\mathbf{l})\in\mathscr{R}\times\mathscr{S},$$
and each summand in Equation~\eqref{dual_C} is  of the form 
$$
\gamma^{s-a_{\mathbf{y},h}}\beta^*_i  \delta_{a_{\mathbf{y},i}}\epsilon_{\mathbf{y}}\mathbf{k}^*_{\mathbf{y}}+\gamma^{s-b_{\mathbf{y}, \mathbf{i}}}\omega_{\mathbf{y}, \mathbf{i}}\epsilon_{\mathbf{y},\mathbf{j}}\mathbf{l}^*_{\mathbf{y},\mathbf{i}},
\hbox{
for some }(\mathbf{k}^*,\mathbf{l}^*)\in\mathscr{R}\times\mathscr{S}.$$ 
Therefore, every $w$ in $D$ satisfies $c \circledast w = 0$ for all $c \in C$, and hence $D \subseteq C^{\perp_{\circledast}}$. Indeed, we have
$$\mathbf{c} \ast \mathbf{w} = \sum_{i=0}^{t-1} \coeffId{\Lambda_i(\mathbf{c}) \Lambda_i(\mathbf{w})} + \chi\left( \sum_{j=t}^{k-1} \coeffId{\overline{\Lambda}_j(\mathbf{c}) \overline{\Lambda}_j(\mathbf{w})} \right),$$
where $\Lambda_i(\mathbf{c}) = \gamma^{a_{\mathbf{x},j}} \beta_j \delta_{a_{\mathbf{x},j}} \Lambda_i(\epsilon_{\mathbf{x}} \mathbf{k}_{\mathbf{x}})$ and $ \Lambda_i(\mathbf{w}) = \gamma^{s - a_{\mathbf{y},h}} \beta_i^* \delta_{a_{\mathbf{y},i}} \Lambda_i(\epsilon_{\mathbf{y}} \mathbf{k}^*_{\mathbf{y}}).$ Thus
$$
\Lambda_i(\mathbf{c}) \Lambda_i(\mathbf{w}) =\gamma^{a_{\mathbf{x},j} + (s - a_{\mathbf{y},i})} \beta_j \beta_i^* \delta_{a_{\mathbf{x},j}} \delta_{a_{\mathbf{y},i}} \Lambda_i(\epsilon_{\mathbf{x}} \mathbf{k}_{\mathbf{x}}) \Lambda_i(\epsilon_{\mathbf{y}} \mathbf{k}^*_{\mathbf{y}}).
$$
If $\mathbf{x} = \mathbf{y}$, then $a_{\mathbf{x},j} + (s - a_{\mathbf{y},i}) = s$, which implies $\gamma^{a_{\mathbf{x},j} + (s - a_{\mathbf{y},i}) }=0$. Thus, the term $\Lambda_i(\mathbf{c}) \Lambda_i(\mathbf{w})=0$.
If $\mathbf{x} \neq \mathbf{y}$, the idempotents satisfy $\epsilon_{\mathbf{x}} \epsilon_{\mathbf{y}} = 0$. So the term $\Lambda_i(\mathbf{c}) \Lambda_i(\mathbf{w})$ vanishes again. Similarly, we can show that $\overline{\Lambda}_j(\mathbf{c})\overline{\Lambda}_j(\mathbf{w})=0$. Therefore, for every $\mathbf{c} \in C$ and $\mathbf{w} \in D$, the trace inner product $\mathbf{c} \circledast \mathbf{w} = \Tr(\mathbf{c} \ast \mathbf{w}) = 0$, which means $D \subseteq C^{\perp_{\circledast}}$.

Finally, taking into account Equation~\eqref{Eq_Iden}, the left $S|R$-additive code $C$ satisfies
$
|C| \cdot |C^{\perp_{\circledast}}| = |\mathscr{S}|.
$
The construction of $D$ by  swapping exponents $a \mapsto s - a$ shows that $D$ has the same number of $R$-coordinates as $C$, so
$
|C| \cdot |D| = |\mathscr{S}|
$
as well. Since we already have $D \subseteq C^{\perp_{\circledast}}$ and both are left additive subcodes of $S$ with the same size, we conclude that $D = C^{\perp_{\circledast}}$.
\end{proof}

\begin{definition}
    The \emph{right trace-annihilator} of a left additive $\mathrm{G}$-code $C$ over $S|R$ is the set
\begin{align}\label{ann}
\Ann^r_{\Tr}(C) \coloneqq \left\{ \mathbf{x} \in \mathscr{S} \mid \Tr(\mathbf{c} \mathbf{x}) = 0 \text{ for all } \mathbf{c} \in C \right\}.
\end{align}
\end{definition}

It is clear that $\Ann^r_{\Tr}(C)$ is a left additive $\mathrm{G}$-code over $S|R$. The following result establishes a connection between the $\circledast$–dual and the right trace-annihilator of a left additive $\mathrm{G}$-code $C$ over the Galois extension $S|R$, expressed through the involutive ring anti-automorphism $\mu$.
\begin{theorem}\label{lemma:trace_dual_annihilator}
    Let $C$ be a left $S|R$-additive $\mathrm{G}$-code. If $S|R$ is a Galois extension, then
    $$
    C^{\perp_{\circledast}} = \mu\left( \Ann^r_{\Tr}(C) \right).
    $$
\end{theorem}

\begin{proof}$\quad$\\
\begin{itemize}
    \item[$\subseteq$)] Let $\mathbf{w} = \sum_{g \in G} w_g g \in \mu\left( \Ann^r_{\Tr}(C) \right)$, and let $ \mathbf{x} = \mu(\mathbf{w})  \in \Ann^r_{\Tr}(C)$. Then  for all $\mathbf{c} = \sum_{g \in G} c_g g \in C$, $\Tr(\mathbf{c}\mathbf{x})= 0$, which implies $ \Tr\left(\coeffId{\mathbf{c}\mathbf{x}}\right) = 0$. Since $S|R$ is a Galois extension, we have $\mathbf{c}\ast \mathbf{w}=\coeffId{\mathbf{c}\mu(\mathbf{w})}$. Therefore
    \[
    \mathbf{c} \circledast \mathbf{w} = \Tr\left( \mathbf{c}\ast \mathbf{w}\right) = \Tr\left(\coeffId{\mathbf{c}\mu(\mathbf{w})}\right) = \Tr\left(\coeffId{\mathbf{c}\mathbf{x}}\right).
    \]
 Thus, $\mathbf{w} \in C^{\perp_{\circledast}}$.
   \item [$\supseteq$)] Let $\mathbf{w} \in C^{\perp_{\circledast}}$, thus $\Tr(\coeffId{\mathbf{c}\mu(\mathbf{w})}) = 0$ for all $\mathbf{c} \in C$.  Since $C$ is a right $\mathscr{R}$-module, $g\mathbf{c} \in C$ for all $g \in G$, and thus
    $
    \coeffId{g^{-1}(\mathbf{c}\mathbf{x})} = 0 \quad \text{for all } g \in G,
    $
    where $\mathbf{x} = \mu(w)$. If we consider $\mathbf{c}\mathbf{x} = \sum_{g \in G} \coeffId{g^{-1}(\mathbf{c}\mathbf{x})} g$, we conclude that $\Tr(\mathbf{c}\mathbf{x}) = 0$ for all $c \in C$. Therefore, $\mathbf{x} \in \Ann^r_{\Tr}(C)$, and consequently $\mathbf{w} = \mu(\mathbf{x}) \in \mu\left( \Ann^r_{\Tr}(C) \right)$. 
    \end{itemize}
\end{proof}

\begin{cor}\label{coro-trace-selforthogonal}
Let $C = \mathscr{R}\mathbf{x}$ be a left $S|R$-additive $\mathrm{G}$-code, where $S|R$ is a Galois extension and $\bf{x} \in \mathscr{R}_\alpha$ satisfies $\Tr(\mathbf{x} \mu(\mathbf{x})) = 0$.  Then $C$ is self-orthogonal with respect to the $\Theta$–Euclidean inner product, i.e.,
$$
C \subseteq C^{\perp_{\circledast}}.
$$
\end{cor}

\begin{proof}
Let $\mathbf{c} = \mathbf{r} \mathbf{x} \in C$ for some $\mathbf{r} \in \mathscr{R}$. For any $\mathbf{r}' \in \mathscr{R}$, consider
$$
\Tr(\mathbf{r}' \mathbf{x}  \mu(\mathbf{c})) = \Tr(\mathbf{r}' \mathbf{x}  \mu(\mathbf{x}) \mu(\mathbf{r})) = \mathbf{r}'  \Tr(\mathbf{x} \mu(\mathbf{x}))\mu(\mathbf{r}).
$$
Since $\Tr(\mathbf{x} \mu(\mathbf{x})) = 0$, it follows that $\Tr(\mathbf{r}' \mathbf{x}  \mu(\mathbf{c})) = 0$ for all $\mathbf{r}' \in \mathscr{R}$.
This means that $\mu(\mathbf{c}) \in \Ann^r_{\Tr}(C)$, and hence $\mathbf{c} \in \mu(\Ann^r_{\Tr}(C)) = C^{\perp_{\circledast}}$ by Theorem \ref{lemma:trace_dual_annihilator}.
\end{proof}

\section{ACP of group codes over finite chain rings}\label{sec:acp}

In this section, $S|R$ denotes an extension of finite chain rings. 
\begin{definition}
    We say that $\{C, D\}$\label{pair} is an \emph{additive complementary pair} (ACP)\label{acp} of left $S|R$-additive codes if $C \oplus D = \mathscr{S}$. 
\end{definition}
When $S = R$,  in \cite{GMS20}   the following lemma was established, which generalizes the case over finite fields previously proved in \cite{BCW}.

\begin{lemma}\label{lem_main} Let $\{C, D\}$ be a pair of two-sided ideals in $\mathscr{R}$. If $C \oplus D = \mathscr{R}$, then 
 $D^{\perp_E} = \mu(C).$
\end{lemma}

From the $\mathscr{R}$-module isomorphism $\Theta \colon \mathscr{S} \longrightarrow (\mathscr{R})^{\ell t} \times (\overline{\mathscr{R}})^{\ell(k-t)}$ defined in \eqref{Lamb}, we immediately obtain the following proposition.

\begin{prop}\label{m10} Let $\{C, D\}$ be a pair of left $S|R$-additive $G$-codes. Then $\{C, D\}$ is ACP if and only if 
  $\{\Theta(C), \Theta(D)\}$ is an LCP\label{lcd} of $\mathscr{R}\overline{\mathscr{R}}$-linear codes of block-length $(\ell t,\ell(k-t))$.
\end{prop}

\begin{lemma}[Proposition 1 in \cite{BDFM25}] Let $R$ be a finite chain ring with maximal ideal generated by $\theta$ and nilpotence index $s$. Let $\overline{R}$ be the ring $R/\theta^rR$ for a fixed $r< s.$ 
If $C$ is an ${R}\overline{{R}}$-linear code of block-length $(\alpha, \beta)$, then there is a unique $s$-tuple of
nonnegative integers $(k_0,\ldots, k_{s-r-1}\,|\,
k_{s-r},\ldots,k_{s-1})$ where $\sum_{i=1}^{s-r-1} k_i<\alpha$ and $\sum_{i=s-r}^{s-1} k_i<\beta$, so-called the \emph{type} of $C,$ such
that $C$ isomorphic to the $R$-module
$$\prod\limits_{\substack{t=0 \\
k_t\neq 0}}^{s-1}\left(R/\langle
\theta^{\,s-t}\rangle\right)^{k_t}.$$ 
\end{lemma}

\begin{definition} With the notation in the previous lemma, an ${R}\overline{{R}}$-linear code of block-length $(\alpha, \beta)$ and type $(k_0,\ldots,
k_{s-r-1}\,|\, k_{s-r},\ldots,k_{s-1})$ is \emph{weakly-free}, if
$k_1=\cdots=k_{s-r-1}=k_{s-r+1}=\cdots=k_{s-1}=0.$ In other words, as $R$ modules $$C\cong\left(R \right)^{k_0}\times \left(R/\langle
\theta^{\,r}\rangle\right)^{k_{s-r}}.$$
\end{definition}

\begin{lemma}[Lemma 2 in \cite{BDFM25}]\label{wea} Let $\{C, D\}$ be an LCP of $R\overline{R}$-linear codes. Then both $C$ and $D$ are also weakly-free
codes.
\end{lemma}

Since the ring $\mathscr{R}$ (the ring $\overline{\mathscr{R}}$) is a  free $R$-module ($\overline{R}$-module), by Lemma~\ref{wea}, we have the following result.

\begin{cor}  If 
$\{C, D\}$ is an LCP of left $S|R$-additive $G$-codes, then $C$ and $D$ are weakly-free.
\end{cor}

\begin{lemma}\label{global-local1} Let $\{C, D\}$  be a pair of left $S|R$-additive $G$-codes. Then  $\{\Theta(C), \Theta(D)\}$ is LCP if and only if 
   $\{\Theta_i(C), \Theta_i(D)\}$ and $\{\overline{\Theta_j}(C), \overline{\Theta_j}(D)\}$ are LCP of left ideals of $\mathscr{R}$, for all $0\leq i < \ell t\leq j <\ell k$.  
\end{lemma}

\begin{proof} From Equation~\eqref{Eq:lam_prod}, we have  
$$\Theta(C) = \left( \prod_{i=0}^{\ell t-1} \Theta_i(C) \right) \times \left( \prod_{j=\ell t}^{\ell k-1} \overline{\Theta}_j(C) \right)$$ and $$\Theta(D) = \left( \prod_{i=0}^{\ell t-1} \Theta_i(D) \right) \times \left( \prod_{j=\ell t}^{\ell k-1} \overline{\Theta}_j(D) \right).$$  
If $\{\Theta(C), \Theta(D)\}$ is LCP, then $ \Theta(C) \oplus \Theta(D) =  \Theta(C\oplus D)= \mathscr{R}^{\ell t}\times \overline{\mathscr{R}}^{\ell(k-1)}. 
$
It follows that $$\Theta_i(C\oplus D)=\Theta_i(C)\oplus \Theta_i(D)= \mathscr{R}\text{ and }\overline{\Theta_j}(C \oplus D)=\overline{\Theta_j}(C)\oplus\overline{\Theta_j}(D)= \overline{\mathscr{R}}$$ for all $0 \leq i < \ell t\leq j< \ell k$.  
Conversely, if $$\Theta_i(C)\oplus \Theta_i(D)= \mathscr{R}\text{ and } \overline{\Theta_j}(C)\oplus\overline{\Theta_j}(D)= \overline{\mathscr{R}}$$ for all $0 \leq i < \ell t\leq j<\ell k$, then  
$$\left(\prod\limits_{i=0}^{\ell t-1} (\Theta_i(C) \oplus \Theta_i(D)) \right)\times\left(\prod\limits_{j=\ell t}^{\ell k-1} (\overline{\Theta_j}(C) \oplus \overline{\Theta_j}(D)) \right)= \mathscr{R}^{\ell t}\times \overline{\mathscr{R}}^{\ell(k-1)}.$$
\end{proof}

\begin{lemma}\label{lem-main-thm2} Suppose that $\{C, D\}$  is a pair of left $S|R$-additive $G$-codes. 
Let $\{\Theta(C), \Theta(D)\}$ be an LCP of two-sided $\mathscr{R}\overline{\mathscr{R}}$-linear codes of block-length $(\ell t,\ell(k-t))$. Then $\wp(\Theta(C))=(\Theta(D))^{\perp_{E}}$, where  {\begin{align}\label{wp}
    \begin{array}{cccc}
  \wp: & \mathscr{R}^{\ell t}\times\overline{\mathscr{R}}^{\ell (k-t)} & \rightarrow & \mathscr{R}^{\ell t}\times\overline{\mathscr{R}}^{\ell (k-t)} \\
     & (u_0,\ldots, u_{\ell t-1}\,|\,\overline{u}_{\ell t}, \ldots, \overline{u}_{\ell k-1}) & \mapsto & (\mu(u_0),\ldots, \mu(u_{\ell t-1})\,|\,\overline{\mu}(\overline{u}_{\ell t}), \ldots, \overline{\mu}(\overline{u}_{\ell k-1})).
\end{array}
\end{align}}%
\end{lemma}

\begin{proof}
Assume that $\{\Theta(C), \Theta(D)\}$ is an LCP of two-sided  $\mathscr{R}\overline{\mathscr{R}}$-linear codes of block-length $(\ell t,\ell(k-t))$. 
By Lemma~\ref{global-local1}, this is equivalent to  $\{\Theta_i(C), \Theta_i(D)\}$ and $\{\overline{\Theta_j}(C), \overline{\Theta_j}(D)\}$ are LCP of two-sided ideals of $\mathscr{R}$, for all $0\leq i < \ell t\leq j <\ell k$.
Now applying  Lemma~\ref{lem_main}, we have $(\Theta_i(D))^{\perp_E} = \mu(\Theta_i(C))$ and $(\overline{\Theta_j}(D))^{\perp_E} = \overline{\mu}(\overline{\Theta_j}(C))$ for all $0\leq i < \ell t\leq j <\ell k$. 
Since
\begin{align*}
   \Theta(C) &= \left( \prod_{i=0}^{\ell t-1} \Theta_i(C) \right) \times \left( \prod_{j=\ell t}^{\ell k-1} \overline{\Theta}_j(C) \right),\\
   \Theta(D) &= \left( \prod_{i=0}^{\ell t-1} \Theta_i(D) \right) \times \left( \prod_{j=\ell t}^{\ell k-1} \overline{\Theta}_j(D) \right),
\end{align*}
we conclude that $\wp(\Theta(C)) = (\Theta(D))^{\perp_E}$.
\end{proof}

\begin{theorem}\label{m11} If $\{C, D\}$ is an ACP of two-sided  $S|R$-additive $G$-codes, then $\mu(C)=D^{\perp_{\circledast}}$.
\end{theorem}

\begin{proof} By Proposition~\ref{m10}, $\{\Theta(C), \Theta(D)\}$ is an LCP of two-sided $\mathscr{R}\overline{\mathscr{R}}$-linear codes.  By Lemma~\ref{lem-main-thm2}, we have 
$
\wp(\Theta(C))=(\Theta(D))^{\perp_{E}}.
$ 
From Theorem~\ref{thm4:main}, we obtain 
$
\Theta(D^{\perp_{\circledast}}) = (\Theta(D))^{\perp_E}.
$
Since 
$
\wp\bigl(\Theta(C)\bigr) = \Theta\bigl(\mu(C)\bigr),
$
it follows $\Theta\bigl(\mu(C)\bigr)=\Theta(D^{\perp_{\circledast}}).$
Since $\Theta$ is a bijective map, we get
$
\mu(C)=D^{\perp_{\circledast}}.
$
\end{proof}

Note that a similar argument to the one in \cite[Theorem~5.10]{BDTM25} leads to the following result.
\begin{lemma}\label{ACP_pi}
   Let $C$ and $D$ be two left $ S | R $-additive codes of length $n$. The pair $ \{C, D\} $ forms an ACP of codes if and only if the pair $\{\pi(C), \pi(D)\}$ also forms an ACP of left $\mathbb F_{q^r}|\mathbb F_{q^d}$-additive $G$-codes. 
\end{lemma}

{ In the serial case $p\nmid |G|$,  the converse of Theorem~\ref{m11} holds.}
\begin{theorem}\label{thm:charpair}
    Let $\{C,D\}$ be a pair of left $S|R$-additive $G$-codes. If  $\mu(C) = D^{\perp_{\circledast}}$ and $p\nmid |G|$, then $\{C, D\}$ is an ACP of codes, and  $C$,  $D$ are two-sided $S|R$-additive $G$-codes. 
\end{theorem}

\begin{proof} From Lemma~\ref{ACP_pi}, it suffices to prove that $\{\pi(C), \pi(D)\}$ forms an ACP of $\mathbb{F}_{p^r}|\mathbb{F}_{p^d}$-additive $G$-codes.   Since $p\nmid |G|$, from Maschke's Theorem {(see for example \cite[Chapter 1]{Webb}), the ambient space $\pi(\mathscr{S})$ is a completely reducible module and hence the lattice of submodules (codes) is a lattice with complements, thus for $\pi(C)$} there is a left $\mathbb{F}_{p^r}|\mathbb{F}_{p^d}$-additive $G$-code $C_1$ such that $\pi(C)\oplus C_1=\pi(\mathscr{S})$. From Theorem \ref{m11}, we have $\mu(\pi(C))=C_1^{\perp_{\circledast}}$. On the other hand, since $\mu(C) = D^{\perp_{\circledast}}$, it follows that $\mu(\pi(C))=\pi(\mu(C)) = \pi(D^{\perp_{\circledast}})=\pi(D)^{\perp_{\circledast}}$. Thus, $C_1=\pi(D)$.  
Finally, both $C$ and $D$ are left $\mathscr{R}$-submodules of $\mathscr{S}$.
Since $\mu$ maps left $\mathscr{R}$-submodules to right $\mathscr{R}$-submodules, $\mu(C)$ is a right $\mathscr{R}$-submodule. The equality $\mu(C)=D^{\perp_{\circledast}}$ implies that
$C$ and $D$ are two-sided, by Lemma \ref{m28}.
\end{proof}
The following example shows that the hypothesis $p\nmid|G|$ is necessary in Theorem~\ref{thm:charpair}.
\begin{example}
 Let $S=R=\mathbb Z_4$ (i.e. the trivial extension), and let $G=\{1,g\}$ be the cyclic group of order $2$. 
Define the left $\mathscr R$-submodules (i.e. left ideals) $C=\mathscr R(1+g)$  and $D=\mathscr R(1-g)$. Both $C$ and $D$ have size $4$ in the ambient ring of size $|\mathscr R|=16$. We have $(1+g)(1-g)=1-g^2=0$, hence $\langle 1+g,1-g\rangle_E=0$. Therefore $C\subseteq D^{\perp_E}$. Since $\mathscr R$ is a finite Frobenius ring, orthogonality satisfies $|D|\cdot|D^{\perp_E}|=|\mathscr R|$. We have $|D^{\perp_E}|=4$, which leads to $D^{\perp_E}=C$. Because $S=R$, the $\circledast$-dual equals the Euclidean dual, so $D^{\perp_{\circledast}}=D^{\perp_E}=C$. Since $\mu$ is the identity map, we get $\mu(C)=D^{\perp_{\circledast}}$.

Each element of $C+D$ can be written as
$$
t(1+g)+s(1-g) = (t+s) + (t-s)g,\qquad t,s\in\mathbb Z_4.
$$

The pair $t=s=2$ provides the same element as $t=s=0$. Hence, $|C+D|=8<16$, and hence $C\oplus D\ \neq\ \mathscr R$. Therefore, the pair of codes $\{C,D\}$ does not form an ACP.   
\end{example}

\begin{prop}\label{prop:AbelianACP} Let $G$ be an Abelian group with $p\nmid |G|$, and let $\{C,D\}$ be an  ACP of left $S|R$-additive $G$-codes whose form is given in Equation~\eqref{deco_C}. If $\underline{\beta}$ is a self-dual basis, then
\begin{eqnarray*} 
  C&=&\sum\limits_{j=0}^{\ell-1}\sum\limits_{\mathbf{x}\in\mathfrak{I}_1^{(q)}}\beta_j  \delta_{{\mathbf{x},j}}\mathscr{K}_{\mathbf{x}} +\sum\limits_{(\mathbf{x}, \mathbf{j})\in\mathfrak{I}_2^{(q)}\times\mathfrak{U}(\flat, \ell)}\omega_{\mathbf{x}, \mathbf{j}}\mathscr{L}_{\mathbf{x}, \mathbf{j}},  \\
   D&=&\sum\limits_{j=0}^{\ell-1}\sum\limits_{\mathbf{x}\in\mathfrak{I}_1^{(q)}}\beta_j  \delta'_{{\mathbf{x},j}}\mathscr{K}_{\mathbf{x}} +\sum\limits_{(\mathbf{x}, \mathbf{j})\in\mathfrak{I}_2^{(q)}\times\mathfrak{U}(\flat, \ell)}\omega'_{\mathbf{x}, \mathbf{j}}\mathscr{L}_{\mathbf{x}, \mathbf{j}},
\end{eqnarray*}
where $\{\delta_{{\mathbf{x},j}};\delta'_{{\mathbf{x},j}};\omega_{\mathbf{x}, \mathbf{j}}, \omega'_{\mathbf{x}, \mathbf{j}}\}\subseteq\{0,1\}$ with $\delta_{{\mathbf{x},j}}+\delta'_{{\mathbf{x},j}}=\omega_{\mathbf{x}, \mathbf{j}}+ \omega'_{\mathbf{x}, \mathbf{j}}=1$.
\end{prop}




\nocite{*}
\bibliographystyle{amsplain}
\bibliography{biblio}

\newpage
\section*{Annex: List of notations}
{\begin{itemize}[leftmargin=1in]
    \item[$R$]: \textit{Finite chain ring  with invariants} $(p,e,d,k_1,t_1)$ of maximal ideal $;\mathfrak{m}_R$ with generator $\gamma_1$ and nilpotency index $s_1$;\;(p.~\pageref{R})
    \item[$S$]: \textit{Finite chain ring  with invariants} $(p,e,d,k_2,t_2)$ of maximal ideal $\mathfrak{m}_S$ with generator $\gamma_2$ with nilpotency index $s_2$;\;(p.~\pageref{S})
        \item [$\mathrm{GR}(p^e, d)$]: \textit{Galois ring} of rank $d$ over $\mathbb{Z}_{p^e}$;\;(p.~\pageref{GR})
    \item[$\Gamma(R)$]: \textit{Teichm\"uller set} of the ring;$R$;\;(p.~\pageref{Ga(R)})
    \item[$\Gamma(R)^*$]: Non-zero elements of the Teichm\"uller set;\;(p.~\pageref{Ga(R)*})
    \item[$f(x)$]: \textit{Monic basic-irreducible polynomial} over $R[x]$ of degree $\ell$;\;(p.~\pageref{f(x)}) 
    \item[$S|R$]: $S$ is the \textit{Galois extension} of $R$, if $S=R[\alpha]$ with $\alpha = x + \langle f(x) \rangle$, where $f(x)$ is a monic basic-irreducible polynomial over $R$ of degree $\ell$;\; (p.~\pageref{R[alpha]})
    \item[$\underline{\alpha}$]: \textit{Ordered polynomial basis} $(1, \alpha, \dots, \alpha^{\ell-1})$ of the free $R$-module $S$;\; (p.~\pageref{basis-alpha})
    \item[$\underline{\beta}$]: \textit{Arbitrary Ordered basis} $(\beta_0,  \dots, \beta_{\ell-1})$ of the free $R$-module $S$;\; (p.~\pageref{basis-beta})
    \item[$\sigma_q$]: $R$-\textit{linear ring automorphism} of $S$ defined by $\sigma_q\left(\sum_{i=0}^{\ell-1} a_i \beta_i\right) = \sum_{i=0}^{\ell-1} a_i \beta_i^q$;\; (p.~\pageref{sigmaq})
    \item[$\Tr$]: \textit{Trace map} of the Galois extension $S | R$, defined as $\Tr = \sum_{i=0}^{\ell-1} \sigma_q^i$;\; (p.~\pageref{trace_map})
    \item[$\mathrm{T}_R^S$]: \textit{Symmetric bilinear form} on $S$ defined as $\mathrm{T}_R^S(x, y) = \Tr(xy)$;(p.~\pageref{trace_bil})
    \item[$\mathrm{M}_{\underline{\beta}}$]: \textit{Gram matrix} of the basis $\underline{\beta}$, defined as $(\mathrm{T}_R^S(\beta_i, \beta_j))_{0 \leq i,j < \ell}$;\; (p.~\pageref{Gram})
    \item[$\underline{\beta}^*$]: \textit{dual-basis} of $\underline{\beta}$, defined as $\underline{\beta}^* = \underline{\beta} \mathrm{M}_{\underline{\beta}}^{-1}$;\; (p.~\pageref{dual-basis})
    \item[$\mathrm{d}_{i,\underline{\beta}}$]: $R$-\textit{module epimorphism} from $S$ to $R$, defined by \\ $\mathrm{d}_{i,\underline{\beta}}(x) = \Tr(x \beta_i^*)$;\; ~\eqref{eq:d_i_alpha} 
    \item[$\mathrm{d}_{\underline{\beta}}$]: $R$-\textit{module isomorphism from} $S$ to $R^\ell$, defined by\\ $\mathrm{d}_{\underline{\beta}}(x) = (\mathrm{d}_{0,\underline{\beta}}(x), \dots, \mathrm{d}_{\ell-1,\underline{\beta}}(x))$;\; ~\eqref{eq:d_alpha}
     \item[$g(x)$]: \textit{Eisenstein polynomial} over $R$ of degree $k$, defined as $g(x) = x^k - \sum_{i=0}^{k-1} a_i x^i$ with $a_i \in \mathfrak{m}_R$ and $a_0 \in \mathfrak{m}_R \setminus \mathfrak{m}_R^2$;\;(p.~\pageref{eisR})
    \item[$g_1(x)$]: Eisenstein polynomial over $\mathrm{GR}(p^e,d)$ of degree $k_1$ ;\; (p.~\pageref{g_1(x)})
    \item[$g_2(x)$]: Eisenstein polynomial over $\mathrm{GR}(p^e,r)$ of degree $k_2$ ; \;(p.~\pageref{g_2(x)}) 
    \item[$S|R$]: $S$ is an \textit{Eisenstein extension} of $R$, if $S=R[\gamma]$ with $\gamma = x + \langle g(x), \gamma_1^{s_1-1} x^t \rangle$, where $g(x)$ is an Eisenstein polynomial over $R$ of degree $k$, and $0 \leq t         \leq k$;\; (p.~\pageref{R[gamma]})
     \item[$\{1, \gamma, \dots, \gamma^{k-1}\}$]:  $R$-independent generating set $\{1, \gamma, \dots, \gamma^{k-1}\}$ of the  $R$-module $S$;\; (p.~\pageref{gamma-basis})
    \item[$S|R$]: $S$ is an extension of finite chain rings with invariants $(p,s_1,\ell ,k,t)$, if $S=R[\alpha,\gamma]$ with $R_\gamma|R$ an Eisenstein extension $R[\gamma]$ of $R$ of block-degree $(k,t)$          and $R_\alpha|R$ Galois extension $R[\alpha]$ of $R$ of degree $\ell$;\; (p.~\pageref{R[alpha,gamma]})
     \item[$(k,t)$]: Block-degree of the Eisenstein extension $S|R$;\; (p.~\pageref{(k,t)})
     \item[$\overline{R}$]: Quotient ring $R / \gamma_1^{s_1-1} R$, a finite chain ring with maximal ideal $\overline{\mathfrak{m}}_R$ and nilpotency index $s_1 - 1$;\; (p.~\pageref{overR})
    \item[$\overline{(\cdot)}$]: Natural projection from $R$ to $\overline{R}$, defined by $x \mapsto \overline{x} = x + \gamma_1^{s_1-1} R$;\; (p.~\pageref{nat-proj})
    \item[$R^t \times \overline{R}^{k-t}$]: Cartesian product with $R$-module structure, elements denoted as $(v_0, \dots, v_{t-1} \mid \overline{v}_t, \dots, \overline{v}_{k-1})$;\; (p.~\pageref{ambian(k,t)})
    \item[$\psi_i$]: $R$-module isomorphism from $\gamma^i R$ to $R$, defined by $\psi_i(\gamma^i x) = x$ for $0 \leq i < t$ defined in  \eqref{psi_i};
    \item[$\overline{\psi}_j$]: $R$-module isomorphism from $\gamma^j R$ to $\overline{R}$, defined by $\overline{\psi}_j(\gamma^j x) = x + \gamma_1^{s_1-1} R$ for $t \leq j < k$ defined in  \eqref{psi_j};
    \item[$\psi$:] $R$-module isomorphism from $\displaystyle\prod_{i=0}^{k-1} \gamma^i R$ to $ R^t \times \overline{R}^{k-t}$ defined in \eqref{psi};
    \item[$\varphi_i$]: $R$-module epimorphism from $S$ to $\gamma^i R$, defined for $0 \leq i < k$;\; (p.~\pageref{varphi-i})
    \item[$\varphi$]: $R$-module isomorphism from $S$ to $\prod_{i=0}^{k-1} \gamma^i R$, defined by $\varphi(x) = (\varphi_0(x), \dots, \varphi_{k-1}(x))$;\; (p.~\pageref{varphi})
    \item[$\xi_i$]: $R$-module epimorphism from $S$ to $R$, defined as $\xi_i = \psi_i \circ \varphi_i$ for $0 \leq i < t$;\; (p.~\pageref{xi-i})
    \item[$\overline{\xi}_j$]: $R$-module epimorphism from $S$ to $\overline{R}$, defined as $\overline{\xi}_j = \overline{\psi}_j \circ \varphi_j$ for $t \leq j < k$;\; (p.~\pageref{xi-j})
    \item[$\overline{\xi}$]: $R$-module isomorphism from $S$ to $R^t\times\overline{R}^{k-t}$ defined in \eqref{xi};
    \item[$\tau$]: Permutation map $(R^t \times \overline{R}^{k-t})^\ell \to R^{t\ell} \times \overline{R}^{(k-t)\ell}$;\; (p.~\pageref{tau})
    \item[$\theta$]: $R$-module isomorphism from $S$ to $R^{t\ell} \times \overline{R}^{(k-t)\ell}$, defined as $$\theta = (\mathrm{d}_{\underline{\beta}}^{\times t} \times \overline{\mathrm{d}}_{\underline{\beta}}^{\times (k-t)}) \circ \xi;\; (p.~\pageref{theta})$$
    \item[$G$]: Finite group of order $n$, written multiplicatively as\\ $G = \{g_1, \dots, g_n\}$ with $1_G=g_1$ as the identity element;\; (p.~\pageref{G})
    \item[$\mathbb{S}_n$]: Symmetric group on $n$ elements;\; (p.~\pageref{Sn})
    \item[\textnormal{$S[G]$}]: Group ring of $G$ over $S$, defined as\\ $S[G] = \left\{ \sum_{i=1}^n a_{g_i} g_i : (a_{g_1}, \dots, a_{g_n}) \in S^n \right\}$;\; (p.~\pageref{SG})
     \item[$\mathscr{S}$]: Notation for the group ring $S[G]$;\;(p.~\pageref{notaSG})
    \item[$\mathscr{R}$]: Notation for the group ring $R[G]$;\;(p.~\pageref{notaRG})
    \item[$1$]: Identity element of $\mathscr{S}$, defined as $1 = 1_S 1_G$;\;(p.~\pageref{1})
    \item[$\mu$]: Involutive ring anti-automorphism $\mu: \mathscr{S} \to \mathscr{S}$, defined by: $$\mu\left(\sum_{i=1}^n x_{g_i} g_i\right) = \sum_{i=1}^n x_{g_i} g_i^{-1},$$ fixing $S$ pointwise defined in \eqref{eq:mu};
    \item[$\pi$]: Ring epimorphism $\pi: \mathscr{S} \to \mathbb{F}_{p^r}[G]$, extending the natural projection $\pi: S \to \mathbb{F}_{p^r}$, defined by $\pi\left(\sum_{g \in G} a_g g\right) = \sum_{g \in G} \pi(a_g) g$;\;(p.~\pageref{pi})
    \item[$\text{Aug}_\mathscr{S}$]: Augmentation map $\text{Aug}_\mathscr{S}: \mathscr{S} \to S$, defined by:\\ $\text{Aug}_\mathscr{S}\left(\sum_{g \in G} a_g g\right) = \sum_{g \in G} a_g$ defined in \eqref{aug}; 
    \item[$\mathfrak{a}_\mathscr{S}$]: Augmentation ideal of $\mathscr{S}$, defined as the kernel of $\text{Aug}_\mathscr{S}$ defined in \eqref{ideal_aug};
    \item[$\coeffId{~\cdot~}$]: $S$-module epimorphism $\coeffId{\cdot}: \mathscr{S} \to S$, defined by $\coeffId{\sum_{g \in G} a_g g} = a_{1_G}$ defined in \eqref{coef};
    \item[$\langle \cdot, \cdot \rangle$]: Symmetric, non-degenerate, $G$-invariant bilinear form on $\mathscr{S}$, defined by $\langle g, h \rangle = \delta_{g,h}$ (Kronecker delta) defined in \eqref{inner};
    \item[$\Phi$]: $S$-module isomorphism $\Phi: S^n \to \mathscr{S}$, defined by $\Phi(a_{g_1}, \dots, a_{g_n}) = \sum_{i=1}^n a_{g_i} g_i$ defined in \eqref{PhiO};
    \item[${}^h(\cdot)$]: Left action of $G$ on $S^n$, defined by ${}^h(a_{g_1}, \dots, a_{g_n}) = (a_{hg_1}, \dots, a_{hg_n})$ for $h \in G$ defined in \eqref{def:left_action};
    \item[$(\cdot)^h$]: Right action of $G$ on $S^n$, defined by $(a_{g_1}, \dots, a_{g_n})^h = (a_{g_1 h}, \dots, a_{g_n h})$ for $h \in G$ defined in \eqref{def:right_action};
    \item[$S|R$-additive code]: Non-empty subset of $S^n$, specifically an $R$-submodule of $S^n$, called an $S|R$-additive code of length $n$; \;(p.~\pageref{defi-add})
    \item[$\mathrm{PAut}(C)$]: Permutation automorphism group of an additive code $C$;\;(p.~\pageref{paut})
    \item[$\flat$]: Number of cyclic factors in the direct product decomposition of the abelian group $G = \langle g_2, \dots, g_\flat \rangle \cong \langle g_2 \rangle \times \cdots \times \langle g_\flat \rangle$;\;(p.~\pageref{decom_G})
    \item[$n_a$]: Order of the cyclic group $\langle g_a \rangle$ for $2 \leq a \leq \flat$;\;(p.~\pageref{n_a})
    \item[$\mathscr{R}_a$]: Group ring $R[\langle g_a \rangle]$ for $2 \leq a \leq \flat$;\;(p.~\pageref{R_a})
    \item[$\mathscr{S}_a$]: Group ring $S[\langle g_a \rangle]$ for $2 \leq a \leq \flat$;\;(p.~\pageref{S_a})
    \item[$\complement_{i,j,a}^{(b)}$]: $b$-cyclotomic coset modulo $n_a$, defined as $\complement_{i,j,a}^{(b)} = \{ i q^j b^h \bmod n_a : h \in \mathbb{Z}^+ \},$ where $b \in \{q, q^\ell\}$, $2 \leq a \leq \flat$, and $0 \leq j < \ell$;\;(p.~\pageref{cyclo})
    \item[$\kappa_{i,j,a}^{(b)}$]: Size of the $b$-cyclotomic coset $\complement_{i,j,a}^{(b)}$, i.e., $\kappa_{i,j,a}^{(b)} = |\complement_{i,j,a}^{(b)}|$;\;(p.~\pageref{Order_cyclo})
    \item[$Z_a^{(q)}$]: Complete set of representatives of all distinct $q$-cyclotomic cosets modulo $n_a$, denoted $Z_a^{(q)} = \{i_0, \dots, i_{v_a}\}$ with $0 = i_0 < i_1 < \cdots < i_{v_a} < n_a$;\;(p.~\pageref{Z_a^q})
    \item[$\mathscr{R}_\alpha$]: Ring $\mathscr{R}[\alpha]$, where $\mathscr{R}$ is a group ring; \;(p.~\pageref{mathscr{R}_alpha})
    \item[$\mathscr{R}_\gamma$]: Ring $\mathscr{R}[\gamma]$, where $\mathscr{R}$ is a group ring; \; (p.~\pageref{mathscr{R}_gamma})
    \item[$\mathrm{D}_{i,\underline{\beta}}$]: $\mathscr{R}_\gamma$-module epimorphism from $\mathscr{S}$ to $\mathscr{R}_\gamma$, defined as $\sum_{g \in G} x_g g \mapsto \sum_{g \in G} \mathrm{d}_{i,\underline{\beta}}(x_g) g$ defined in \eqref{eq:D_i_alpha};
    \item[$\mathrm{D}_{\underline{\beta}}$]: $\mathscr{R}_\gamma$-module isomorphism from $\mathscr{S}$ to $(\mathscr{R}_\gamma)^\ell$, defined as $x \mapsto (\mathrm{D}_{0,\underline{\beta}}(x), \dots, \mathrm{D}_{\ell-1,\underline{\beta}}(x))$ defined in \eqref{eq:D_alpha};
    \item[$\Psi$]: Map from $\prod_{i=0}^{k-1} \gamma^i \mathscr{R}_\alpha$ to $(\mathscr{R}_\alpha)^t \times (\overline{\mathscr{R}}_{\overline{\alpha}})^{k-t}$, defined as\\ $(x_0, \dots, x_{k-1}) \mapsto (\Psi_0(x_0), \dots, \Psi_{t-1}(x_{t-1}) \mid \overline{\Psi}_t(x_t), \dots, \overline{\Psi}_{k-1}(x_{k-1}))$ defined in \eqref{Psi1};
    \item[$\Psi_i$]: Map from $\gamma^i \mathscr{R}_\alpha$ to $\mathscr{R}_\alpha$, defined as $\gamma^i x \mapsto x$ for $0 \leq i < t$ defined in \eqref{psi_i};
    \item[$\overline{\Psi}_j$]: Map from $\gamma^j \mathscr{R}_\alpha$ to $\overline{\mathscr{R}}_{\overline{\alpha}}$, defined as $\gamma^j x \mapsto x + \gamma_1^{s_1-1}\mathscr{R}$ for $t \leq j < k$ defined in \eqref{psi_j};
    \item[$\phi_i$]: $\mathscr{R}_\alpha$-module epimorphism from $\mathscr{S}$ to $\gamma^i \mathscr{R}_\alpha$, defined as $x \mapsto \phi_i(x)$ for $0 \leq i < k$ defined in \eqref{phi_i}; 
    \item[$\phi$]: $\mathscr{R}$-module isomorphism from $\mathscr{S}$ to $\prod_{i=0}^{k-1} \gamma^i \mathscr{R}_\alpha$, defined as $x \mapsto (\phi_0(x), \dots, \phi_{k-1}(x))$ defined in \eqref{phi};
    \item[$\Lambda$]: $\mathscr{R}$-module isomorphism from $\mathscr{S}$ to $(\mathscr{R}_\alpha)^t \times (\overline{\mathscr{R}}_{\overline{\alpha}})^{k-t}$, defined as $\Lambda = \Psi \circ \phi$ defined in \eqref{upsilon};
    \item[$\Lambda_i$]: $\mathscr{R}$-module epimorphism from $\mathscr{S}$ to $\mathscr{R}_\alpha$, defined as $\Lambda_i = \Psi_i \circ \phi_i$ for $0 \leq i < t$ defined in \eqref{upsilons};
    \item[$\overline{\Lambda}_j$]: $\mathscr{R}$-module epimorphism from $\mathscr{S}$ to $\overline{\mathscr{R}}_{\overline{\alpha}}$, defined as $\overline{\Lambda}_j = \overline{\Psi}_j \circ \phi_j$ for $t \leq j < k$ defined in \eqref{upsilons};
    \item[$\Theta$]: $\mathscr{R}$-module isomorphism from $\mathscr{S}$ to $(\mathscr{R}_\alpha)^{\ell t} \times (\overline{\mathscr{R}}_{\overline{\alpha}})^{\ell(k-t)}$, defined as $\Theta = \left(\mathrm{D}_{\underline{\beta}}^{\times t} \times \overline{\mathrm{D}}_{\underline{\beta}}^{\times (k-t)}\right) \circ \Lambda$, see \eqref{Lamb};
    \item[$\Theta_i$]: $\mathscr{R}$-module epimorphism from $\mathscr{S}$ to $\mathscr{R}$ for $0 \leq i < \ell t$;defined in \eqref{lambs};
    \item[$\overline{\Theta}_j$]: $\mathscr{R}$-module epimorphism from $\mathscr{S}$ to $\overline{\mathscr{R}}$ for $\ell t \leq j < \ell k$ defined in \eqref{lambs};
    \item[$\ast$]: Euclidean inner product on the $R_{\alpha}$-module $\mathscr{S}$, defined as\\ $\mathbf{v} \ast \mathbf{w} = \sum_{i=0}^{t-1} \coeffId{\Lambda_i(\mathbf{v})\Lambda_i(\mathbf{w})} + \chi\left(\sum_{j=t}^{k-1} \coeffId{\overline{\Lambda}_j(\mathbf{v})\overline{\Lambda}_j(\mathbf{w})}\right)$ defined in \eqref{eq:upsi_inner};
    \item[$\chi$]: $R$-module isomorphism from $\overline{R}_{\overline{\alpha}}$ to $R_{\alpha}\mathfrak{m}_R$, defined as \\ $\sum_{i=0}^{s-2} x_{i} \overline{\gamma_1}^i \mapsto \gamma_1 \sum_{i=0}^{s-2}  \iota(x_{i}) \gamma_1^i$ defined in \eqref{chi};
    \item[$\iota$]: Map from $\Gamma(\overline{R})$ to $\Gamma(R)$, with restriction to $\Gamma(\overline{R})^*$ being a group isomorphism and $\iota(0) = 0$;\;(p.~\pageref{iota})
    \item[$\circledast$]:   inner product on $\mathscr{S}$, defined as $\mathbf{v} \circledast \mathbf{w} = \Tr(\mathbf{v} \ast \mathbf{w})$ defined in \eqref{eq:inner};
\item[$C^{\perp_{\circledast}}$]: ${\circledast}$-dual of a left $S|R$-additive $G$-code $C$, defined as \\ $C^{\perp_{\circledast}} = \{ \mathbf{w} \in \mathscr{S} : \mathbf{c} \circledast \mathbf{w} = 0 \text{ for all } \mathbf{c} \in C \}$ defined in \eqref{dual};
    \item[\textnormal{$[\cdot,\cdot]_E$}]: Euclidean inner product on $\mathscr{R}^{t\ell} \times \overline{\mathscr{R}}^{(k-t)\ell}$, related to $\circledast$ via $[\Theta(\mathbf{v}), \Theta(\mathbf{w})]_E = \sum_{g \in G} (g^{-1} \mathbf{v} \circledast \mathbf{w})$ defined in \eqref{[]_E};
    \item[$\Ann^r_{\Tr}(C)$]: Right trace-annihilator of a left additive $G$-code $C$ over $S|R$, defined as $\Ann^r_{\Tr}(C) = \{ x \in \mathscr{S} : \Tr(c x) = 0 \text{ for all } c \in C \}$  defined in \eqref{ann}. 
    \item[$\{C, D\}$]:  A pair of left $S|R$-additive $G$-codes;(p.~\pageref{pair})
    \item[ACP]: Additive Complementary Pair, a pair $\{C, D\}$ of left $S|R$-additive $G$-codes such that $C \oplus D = \mathscr{S}$;(p.~\pageref{acp})
    \item[LCP]: Linear Complementary Pair, a pair $\{\Theta(C), \Theta(D)\}$ of $\mathscr{R}\overline{\mathscr{R}}$-linear codes such that $\Theta(C) \oplus \Theta(D) = \mathscr{R}^{\ell t} \times \overline{\mathscr{R}}^{\ell(k-t)}$;(p.~\pageref{lcd})
    \item[$\overline{\mu}$]: Induced map on $\overline{\mathscr{R}}$ defined in \eqref{eq:mu};
    \item[$\wp$]: Map from $\mathscr{R}^{\ell t} \times \overline{\mathscr{R}}^{\ell(k-t)}$ to itself, defined by:\\{\footnotesize $\wp(u_0, \dots, u_{\ell t-1} \mid \overline{u}_{\ell t}, \dots, \overline{u}_{\ell k-1}) = (\mu(u_0), \dots, \mu(u_{\ell t-1}) \mid \overline{\mu}(\overline{u}_{\ell t}), \dots, \overline{\mu}(\overline{u}_{\ell k-1}))$} defined in \eqref{wp}.   
\end{itemize}}
 
\end{document}